\UseRawInputEncoding
\documentclass[aip,jcp,reprint,twocolumn,superscriptaddress,showpacs,floatfix]{revtex4-1}

\usepackage{epsfig}
\usepackage{mathrsfs}
\usepackage{graphicx}
\usepackage{amssymb}
\usepackage{dcolumn}
\usepackage{bm}
\usepackage{gensymb}
\usepackage{physics}
\usepackage{setspace}
\usepackage{float}
\usepackage{longtable}
\usepackage{multirow}
\usepackage{color}
\definecolor{red}{rgb}{1,0,0}

\begin{document}

\title{High-Level Coupled-Cluster Energetics by Merging Moment 
Expansions with Selected Configuration Interaction} 

\author{Karthik Gururangan}
\affiliation{Department of Chemistry, Michigan State University, East Lansing, Michigan 48824, USA}

\author{J. Emiliano Deustua} 
\thanks{Present address: Division of Chemistry and Chemical Engineering,
California Institute of Technology, Pasadena, California 91125, USA.}
\affiliation{Department of Chemistry, Michigan State University, East Lansing, Michigan 48824, USA}

\author{Jun Shen}
\affiliation{Department of Chemistry, Michigan State University, East Lansing, Michigan 48824, USA}

\author{Piotr Piecuch}
\thanks{Corresponding author}
\email[e-mail: ]{piecuch@chemistry.msu.edu.}
\affiliation{Department of Chemistry, Michigan State University, East Lansing, Michigan 48824, USA}
\affiliation{Department of Physics and Astronomy, Michigan State University, East Lansing, Michigan 48824, USA}

\date{\today}

\begin{abstract}
Inspired by our earlier semi-stochastic work aimed at converging high-level
coupled-cluster (CC) energetics
[J. E. Deustua, J. Shen,  and P. Piecuch, Phys. Rev. Lett. \textbf{119}, 223003 (2017);
J. Chem. Phys. \textbf{154}, 124103 (2021)], we propose a novel form of the CC($P$;$Q$)
theory in which the stochastic Quantum Monte Carlo propagations,
used to identify dominant higher--than--doubly excited determinants, are replaced by the selected
configuration interaction (CI) approach using the CIPSI algorithm.
The advantages of the resulting CIPSI-driven CC($P$;$Q$) methodology are illustrated by a few
molecular examples, including the dissociation of $\mathrm{F_2}$ and the automerization
of cyclobutadiene, where we recover the electronic energies corresponding to the CC calculations with
a full treatment of singles, doubles, and triples based on the information extracted from
compact CI wave functions originating from relatively inexpensive Hamiltonian diagonalizations.
\end{abstract}

\maketitle



\section{Introduction}
\label{sec1}

One of the key objectives of quantum chemistry is to obtain accurate energetics of molecular systems in a
computationally efficient manner. Among the various post-Hartree--Fock (post-HF) theories,
the size extensive
approaches derived from the exponential ansatz \cite{Hubbard:1957,Hugenholtz:1957}
of coupled-cluster (CC) theory\cite{Coester:1958,Coester:1960,cizek1,cizek2,cizek4} are among the best techniques
to accomplish this task.\cite{paldus-li,bartlett-musial2007}
We recall that the CC wave function for an $N$-electron system is defined as
\begin{equation}
	|\Psi\rangle = e^T|\Phi\rangle,
	\label{eq:ccansatz}
\end{equation}
where $\ket{\Phi}$ is the reference (usually, HF) determinant and 
\begin{equation}
	T = \sum_{n=1}^{N}T_n
	\label{eq:clusterop}
\end{equation}
is the cluster operator, with $T_{n}$
representing its $n$-body ($n$-particle--$n$-hole) component. In practice, one truncates
the cluster operator $T$ at a given many-body rank to define the standard CC hierarchy of
approximations. The most basic and most practical one, obtained when $T$ is truncated at $T_2$,
which has computational steps that scale as $\mathcal{O}(\mathcal{N}^6)$ with the system size $\mathcal{N}$,
is the CC method with singles and doubles (CCSD).\cite{ccsd,ccsd2} The next two levels, namely, the CC
approach with singles, doubles, and triples, abbreviated as CCSDT,\cite{ccfullt,ccfullt2,ch2-bartlett2}
which interests us in this study most, obtained when $T$ is truncated at $T_3$,
and the CC method with singles, doubles, triples, and quadruples,
abbreviated as CCSDTQ,\cite{ccsdtq0,ccsdtq2,ccsdtq3} in which $T$ is truncated at $T_4$,
involve the $\mathcal{O}(\mathcal{N}^8)$ and $\mathcal{O}(\mathcal{N}^{10})$ steps, respectively.
It is well established that in the majority of chemical applications,
including molecules near equilibrium geometries, bond dissociations involving
smaller numbers of strongly correlated electrons, noncovalent interactions, and
photochemistry, the conventional CCSD, CCSDT, CCSDTQ, etc.\ hierarchy and its equation-of-motion (EOM)
\cite{emrich,eomcc1,eomcc3,eomccsdt1,eomccsdt2,eomccsdt3,kallaygauss,hirata1} and linear-response
\cite{monk,monk2,mukherjee_lrcc,sekino-rjb-1984,lrcc3,lrcc4,jorgensen,kondo-1995,kondo-1996}
extensions rapidly approach the exact, full configuration interaction (FCI) limit, so that
by the time one reaches the CCSDT or CCSDTQ levels, the results are usually converged with
respect to the relevant many-electron correlation effects,\cite{bartlett-musial2007} but the
$\mathcal{O}(\mathcal{N}^8)$ computational steps of CCSDT or the $\mathcal{O}(\mathcal{N}^{10})$
steps of CCSDTQ render the usage of such methods unfeasible for most problems of interest.
Thus, one of the main activities in the CC development work has been the design of high-fidelity approximations
to CCSDT and CCSDTQ, capable of reducing the above costs, while being more robust than
perturbative approaches of the CCSD(T)\cite{ccsdpt} type, which fail in more multi-reference situations.
\cite{paldus-li,bartlett-musial2007,irpc,PP:TCA,piecuch-qtp}

To that end, our group has recently developed the semi-stochastic CC($P$;$Q$) formalism,
\cite{stochastic-ccpq-prl-2017,stochastic-ccpq-jcp-2021,stochastic-ccpq-molphys-2020}
a novel methodology that can efficiently converge the energetics of high-level CC calculations, 
such as CCSDT, CCSDTQ, and EOMCCSDT,\cite{eomccsdt1,eomccsdt2,eomccsdt3} by
combining the deterministic CC($P$;$Q$) framework
\cite{jspp-chemphys2012,jspp-jcp2012,jspp-jctc2012,nbjspp-molphys2017,ccpq-be2-jpca-2018} with 
the stochastic Quantum Monte Carlo (QMC) wave function propagations in the many-electron Hilbert space
defining the CIQMC\cite{Booth2009,Cleland2010,fciqmc-uga-2019,ghanem_alavi_fciqmc_jcp_2019,ghanem_alavi_fciqmc_2020}
and CC Monte Carlo (CCMC)\cite{Thom2010,Franklin2016,Spencer2016,Scott2017} approaches. The semi-stochastic
CC($P$;$Q$) methodology of Refs.\
\onlinecite{stochastic-ccpq-prl-2017,stochastic-ccpq-jcp-2021,stochastic-ccpq-molphys-2020}
leverages the fact, recognized long time ago in the context of active-space CC considerations (cf. Ref.\
\onlinecite{piecuch-qtp} for a review), that higher-order cluster operators, such as $T_3$ and $T_4$,
and their counterparts utilized in EOMCC are usually relatively sparse.
In the semi-stochastic CC($P$;$Q$) approaches, the dominant
higher--than--doubly excited cluster/excitation amplitudes relevant to the parent CC/EOMCC
theory of interest are automatically selected using stochastic CIQMC or CCMC wave function propagations
that provide lists of Slater determinants for the initial CC($P$)
\cite{stochastic-ccpq-prl-2017,stochastic-ccpq-jcp-2021} or EOMCC($P$)
\cite{eomccp-jcp-2019,stochastic-ccpq-molphys-2020} calculations, which are
subsequently corrected using the biorthogonal moment expansions adopted in the CC($P$;$Q$) formalism
to capture the remaining correlations. The semi-stochastic CC($P$;$Q$) methods have demonstrated their ability to
rapidly converge the CCSDT,\cite{stochastic-ccpq-prl-2017,stochastic-ccpq-jcp-2021,%
stochastic-ccpq-molphys-2020} CCSDTQ,\cite{stochastic-ccpq-jcp-2021} and EOMCCSDT\cite{stochastic-ccpq-molphys-2020}
energetics out of the early stages of the underlying CIQMC or CCMC propagations,
with minimal reliance on user- and system-dependent inputs.

Encouraged by the above findings, in this study we explore the use of selected CI as 
an alternative provider of the lists of the leading higher--than--doubly excited determinants 
needed to drive the CC($P$;$Q$) computations. The selected CI schemes, which date back to
the pioneering efforts in the late 1960s and early 1970s,\cite{sci_1,sci_2,sci_3,sci_4}
have recently regained significant interest, as their modern implementations
have demonstrated the ability to capture the bulk of many-electron correlation effects 
in a computationally efficient manner using a conceptually straightforward linear 
wave function ansatz.
\cite{adaptive_ci_1,adaptive_ci_2,asci_1,asci_2,ici_1,ici_2,shci_1,shci_2,shci_3,
cipsi_1,cipsi_2,cipsi_benzene,eriksen-et-al-jpcl-2020}
The selected CI model adopted in the CC($P$;$Q$) considerations
reported in this work is the CI method using 
perturbative selection made iteratively (CIPSI),\cite{sci_3} as 
recently reformulated and further developed in Refs.\ \onlinecite{cipsi_1,cipsi_2}.


\section{Theory and Algorithmic Details}
\label{sec2}

We begin by reviewing the key elements of the ground-state CC($P$;$Q$) formalism
relevant to this work. Each CC($P$;$Q$) calculation starts by
identifying two disjoint subspaces of the $N$-electron Hilbert space, the $P$ space designated
as $\mathscr{H}^{(P)}$ and the $Q$ space denoted as $\mathscr{H}^{(Q)}$. The former space is spanned by 
the excited determinants $|\Phi_K\rangle = E_K|\Phi\rangle$, where $E_K$ is the
elementary particle--hole excitation operator generating $|\Phi_K\rangle$ from $|\Phi\rangle$,
which together with $|\Phi\rangle$ dominate the ground-state wave function, whereas
the determinants in $\mathscr{H}^{(Q)}$
are used to construct the correction
due to the correlation effects the CC calculations in the $P$ space do not
describe.
Once the $P$ and $Q$ spaces are defined, we solve
the CC amplitude equations
\begin{equation}
        \mathfrak{M}_K(P) = 0, \;\;
        |\Phi_K\rangle \in \mathscr{H}^{(P)},
        \label{eq:cceqns}
\end{equation}
where
\begin{equation}
        \mathfrak{M}_K(P) = \langle \Phi_K | \overline{H}^{(P)}|\Phi\rangle ,
        \label{eq:moment}
\end{equation}
with
\begin{equation}
	\overline{H}^{(P)} = e^{-T^{(P)}}He^{T^{(P)}} ,
	\label{eq:hbar_p}
\end{equation}
are moments of the CC equations,\cite{moments,leszcz,ren1}
to obtain the approximate form of the cluster operator in the $P$ space,
\begin{equation}
	T^{(P)} = \sum_{\ket{\Phi_K} \in \mathscr{H}^{(P)}} t_K E_K ,
	\label{eq:clusterop_p}
\end{equation}
and the corresponding ground-state energy
\begin{equation}
	E^{(P)} = \langle \Phi | \overline{H}^{(P)} | \Phi \rangle,
	\label{eq:ccenergy}
\end{equation}
and calculate the noniterative correction $\delta(P;Q)$ to determine the
final CC($P$;$Q$) energy as
\begin{equation}
        E^{(P+Q)} = E^{(P)} + \delta(P;Q).
        \label{eq:ccpq_energy}
\end{equation}
The correction $\delta(P;Q)$ to the energy $E^{(P)}$ obtained in the
$P$-space CC [CC($P$)] calculations is given by\cite{jspp-chemphys2012,jspp-jcp2012}
\begin{equation}
        \delta(P;Q) = \sum_{\ket{\Phi_K} \in \mathscr{H}^{(Q)}} \ell_K(P) \: \mathfrak{M}_K(P),
\label{eq:mmcorrection}
\end{equation}
where $\mathfrak{M}_K(P)$ is defined by Eq. (\ref{eq:moment}) and
\begin{equation}
	\ell_K(P) = \langle \Phi| (1 + \Lambda^{(P)}) \overline{H}^{(P)} | \Phi_K\rangle 
	/ D_K^{(P)} ,
	\label{eq:mmleftamp}
\end{equation}
with
\begin{equation}
	D_K^{(P)} = E^{(P)} - \langle \Phi_K | \overline{H}^{(P)} | \Phi_K \rangle .
	\label{eq:mmdenom}
\end{equation}
The $\Lambda^{(P)}$ operator entering Eq. (\ref{eq:mmleftamp}), given by
\begin{equation}
	\Lambda^{(P)} =
	\sum_{\ket{\Phi_K}\in\mathscr{H}^{(P)}} \lambda_K (E_K)^\dag
	\label{eq:leftccop}
\end{equation}
and obtained by solving the linear system
\begin{equation}
        \bra{\Phi}(1 + \Lambda^{(P)})\overline{H}^{(P)}\ket{\Phi_K} = E^{(P)} \lambda_{K} , \;\;
        \ket{\Phi_K}\in\mathscr{H}^{(P)} ,
        \label{eq:leftcceqns}
\end{equation}
is the hole--particle deexcitation operator defining the bra state
$\langle \tilde{\Psi} ^{(P)}| = \langle \Phi | (1 + \Lambda^{(P)}) e^{-T^{(P)}}$ 
associated with the CC($P$) ket state $|\Psi^{(P)} \rangle = e^{T^{(P)}}|\Phi\rangle$.

The CC($P$;$Q$) formalism includes the completely renormalized CC methods, such as
CR-CC(2,3),\cite{crccl_jcp,crccl_cpl,crccl_molphys,crccl_jpc} which works better
than CCSD(T) in bond breaking situations, but its main advantage is the freedom
to make unconventional choices of the $P$ and $Q$ spaces, allowing one to relax the
lower-order $T_1$ and $T_2$ clusters in the presence of their higher-order counterparts, such as the
leading $T_{3}$ contributions, which the CCSD(T), CR-CC(2,3), and other triples corrections to CCSD
cannot do. One can use active orbitals to identify the leading
higher--than--doubly excited determinants for the inclusion in the $P$ space
used in the CC($P$) calculations, employing the
$\delta(P;Q)$ corrections to capture the remaining correlations of interest, as in the CC(t;3)
and similar approaches,
\cite{jspp-chemphys2012,jspp-jcp2012,jspp-jctc2012,nbjspp-molphys2017,ccpq-be2-jpca-2018}
or adopt a more black-box semi-stochastic CC($P$;$Q$) framework, in which
the selection of the dominant higher--than--doubly excited determinants entering the $P$ space
is automated with the help of CIQMC or CCMC propagations.
\cite{stochastic-ccpq-prl-2017,stochastic-ccpq-jcp-2021,stochastic-ccpq-molphys-2020}
In this article, we propose an alternative to the semi-stochastic CC($P$;$Q$) methodology,
in which we use the information extracted from the CIPSI runs to populate the $P$ spaces employed
in the CC($P$) calculations preceding the determination of the $\delta(P;Q)$ corrections.

We recall that the CIPSI approach, originally proposed in Ref.\ \onlinecite{sci_3} and further developed
in Refs.\ \onlinecite{cipsi_1,cipsi_2},
seeks to construct an approximation to the FCI wave function by a series of Hamiltonian
diagonalizations in increasingly large, iteratively defined, subspaces of the many-electron
Hilbert space, designated as $\mathcal{V}_{\text{int}}^{(k)}$, where $k = 0, 1, 2,\ldots$ enumerates
the consecutive CIPSI iterations. The initial subspace $\mathcal{V}_{\text{int}}^{(0)}$
can be one-dimensional, if the CIPSI calculations are started from a single determinant,
such as the restricted HF (RHF) wave function used throughout this work as a reference,
or multi-dimensional, if one prefers to start from a multi-determinantal state generated
in some preliminary truncated CI computation, and the remaining subspaces are constructed
via a recursive process in which $\mathcal{V}_{\text{int}}^{(k+1)}$ is obtained by
augmenting $\mathcal{V}_{\text{int}}^{(k)}$ with a subset of the leading singly and doubly
excited determinants out of $\mathcal{V}_{\text{int}}^{(k)}$ identified with the help of the
many-body perturbation theory (MBPT). Thus, if
$|\Psi^{(\text{CIPSI})}_k\rangle = \sum_{|\Phi_I\rangle \in \mathcal{V}_{\text{int}}^{(k)}} c_{I} |\Phi_I\rangle$
is a CI wave function obtained by diagonalizing the Hamiltonian in the current subspace
$\mathcal{V}_{\text{int}}^{(k)}$ and $E_{\text{var},k}$ is the corresponding energy, and
if $\mathcal{V}_{\text{ext}}^{(k)}$ is the space of all singly and doubly excited
determinants out of $|\Psi^{(\text{CIPSI})}_k\rangle$, for each determinant
$|\Phi_\alpha\rangle \in \mathcal{V}_{\text{ext}}^{(k)}$ we evaluate the second-order MBPT
correction
$e_{\alpha,k}^{(2)} = | \langle \Phi_\alpha | H | \Psi_k^{(\text{CIPSI})} \rangle |^2 /
(E_{\text{var},k} - \langle \Phi_\alpha | H | \Phi_\alpha \rangle)$
and use the resulting $e_{\alpha,k}^{(2)}$ values to decide which determinants from
$\mathcal{V}_{\text{ext}}^{(k)}$ should be added to the determinants $|\Phi_I\rangle$ already
in $\mathcal{V}_{\text{int}}^{(k)}$ to construct the next diagonalization space
$\mathcal{V}_{\text{int}}^{(k+1)}$.
We can also use the $e_{\alpha,k}^{(2)}$ values to calculate the perturbatively
corrected CIPSI energies $E_{\text{var},k} + \Delta E^{(2)}_k$, where
$\Delta E^{(2)}_k = \sum_{|\Phi_\alpha\rangle \in \mathcal{V}_{\text{ext}}^{(k)}} e_{\alpha,k}^{(2)}$, and,
after further manipulations, their $E_{\text{var},k} + \Delta E_{\text{r},k}^{(2)}$ counterparts, in which
$\Delta E^{(2)}_k$ is replaced by its renormalized $\Delta E_{\text{r},k}^{(2)}$ form
introduced in Ref.\ \onlinecite{cipsi_2}.

In the modern implementation of CIPSI, formulated in Refs. \onlinecite{cipsi_1,cipsi_2} and available in
the Quantum Package 2.0 software,\cite{cipsi_2} which we used in the present study, the process of
enlarging $\mathcal{V}_{\text{int}}^{(k)}$ to generate $\mathcal{V}_{\text{int}}^{(k+1)}$ is executed in the
following manner. First, prior to examining the $e_{\alpha,k}^{(2)}$ corrections, one stochastically
samples the most important singly and doubly excited determinants out of $|\Psi^{(\text{CIPSI})}_k\rangle$,
so that not all singles and doubles from $\mathcal{V}_{\text{int}}^{(k)}$ are included in the accompanying
$\mathcal{V}_{\text{ext}}^{(k)}$ space, only the sampled ones. Next, one arranges the sampled determinants
$|\Phi_\alpha\rangle \in \mathcal{V}_{\text{ext}}^{(k)}$ in descending
order according to their $|e^{(2)}_{\alpha,k}|$ values.
The process of enlarging the current subspace $\mathcal{V}_{\text{int}}^{(k)}$ to construct
the $\mathcal{V}_{\text{int}}^{(k+1)}$ space for the subsequent Hamiltonian diagonalization, which
starts from the determinants $|\Phi_\alpha\rangle$ characterized by the largest $|e^{(2)}_{\alpha,k}|$
contributions, moving toward those that have smaller $|e^{(2)}_{\alpha,k}|$ values, continues
until the number of determinants in $\mathcal{V}_{\text{int}}^{(k+1)}$ exceeds the dimension of
$\mathcal{V}_{\text{int}}^{(k)}$ multiplied by a user-defined factor $f > 1$. In this study, we used
$f=2$, which is the default in Quantum Package 2.0 (we will examine other choices of $f$ in the future).
In practice, a typical dimension of $\mathcal{V}_{\text{int}}^{(k+1)}$, including each of the final diagonalization
spaces used to generate lists of higher--than--doubly excited determinants for the CC($P$) calculations reported in
this work, is slightly larger than $f$ times the dimension of $\mathcal{V}_{\text{int}}^{(k)}$, since the
CIPSI algorithm adds extra determinants to $\mathcal{V}_{\text{int}}^{(k+1)}$ to ensure that the resulting
$|\Psi_{k+1}^{(\text{CIPSI})}\rangle$ wave function is an eigenstate of the total spin $S^2$ and $S_z$ operators.
The final wave function $|\Psi^{(\text{CIPSI})}\rangle$ of a given
CIPSI run and the associated variational ($E_{\text{var}}$) and perturbatively corrected
[$E_{\text{var}} + \Delta E^{(2)}$ and $E_{\text{var}} + \Delta E_{\text{r}}^{(2)}$] energies
are obtained by terminating the above procedure in one of the following two ways:
(i) stopping at the first iteration $k$ for which the second-order MBPT correction $|\Delta E^{(2)}_k|$ falls
below a user-defined threshold $\eta$, indicating that the CIPSI wave function is within a tolerable distance from FCI,
or (ii) stopping at the first iteration $k$ for which the number of determinants in the corresponding
$\mathcal{V}_{\text{int}}^{(k)}$ space exceeds a user-defined input parameter $N_{\text{det(in)}}$.
Since our main objective is to employ the CIPSI-driven CC($P$;$Q$) algorithm to accurately reproduce
the high-level CC rather than FCI energetics, without having to converge the underlying CIPSI sequence,
we chose the latter option, which we enforced by using $\eta = 10^{-6}$ hartree. As a result of setting
the input parameter $f$ at 2, the sizes of the final wave functions $|\Psi^{(\text{CIPSI})}\rangle$
produced by our CIPSI runs, denoted as $N_{\text{det(out)}}$, were always between $N_{\text{det(in)}}$
and $2 N_{\text{det(in)}}$.

Having discussed the key ingredients of the CC($P$;$Q$) and CIPSI methodologies relevant to this work,
we proceed to the description of the CIPSI-driven CC($P$;$Q$) algorithm, which consists of the
following steps:
\begin{itemize}
\item[1.]
Given a reference state $|\Phi\rangle$, which in all of the calculations reported
in this article was the RHF determinant, choose an input parameter $N_{\text{det(in)}}$,
used to terminate the CIPSI wave function growth, and execute a CIPSI run
to obtain the final state $|\Psi^{(\text{CIPSI})}\rangle$ spanned by $N_{\text{det(out)}}$ determinants.
\item[2.]
Extract a list of higher--than--doubly excited determinants relevant to the desired CC theory
level from $|\Psi^{(\text{CIPSI})}\rangle$ to define the $P$ space. If the goal is to
converge the CCSDT energetics, the $P$ space consists of all singly and doubly excited determinants
plus the triply excited determinants contained in $|\Psi^{(\text{CIPSI})}\rangle$. To recover the
CCSDTQ energetics, quadruply excited determinants contributing to $|\Psi^{(\text{CIPSI})}\rangle$
are included in the $P$ space as well.
\item[3.]
Solve the nonlinear CC($P$) system, Eq. (\ref{eq:cceqns}), and the associated linear system given by
Eq. (\ref{eq:leftcceqns}), where $E^{(P)}$ is defined by Eq. (\ref{eq:ccenergy}), in the $P$ space determined
in Step 2 to obtain the cluster operator $T^{(P)}$ and the deexcitation operator ${\Lambda}^{(P)}$.
If the target approach is CCSDT, define $T^{(P)} = T_1 + T_2 + T_3^{(\text{CIPSI})}$ 
and $\Lambda^{(P)} = \Lambda_1 + \Lambda_2 + \Lambda_3^{(\text{CIPSI})}$, 
where the list of triples entering $T_3^{(\text{CIPSI})}$ and $\Lambda_3^{(\text{CIPSI})}$
is identical to that extracted from $|\Psi^{(\text{CIPSI})}\rangle$ in Step 2.
If the goal is to converge the CCSDTQ energetics, define 
$T^{(P)} = T_1 + T_2 + T_3^{(\text{CIPSI})} + T_4^{(\text{CIPSI})}$ and
$\Lambda^{(P)} = \Lambda_1 + \Lambda_2 + \Lambda_3^{(\text{CIPSI})} + \Lambda_4^{(\text{CIPSI})}$,
where the list of triples entering $T_3^{(\text{CIPSI})}$ and $\Lambda_3^{(\text{CIPSI})}$
and the list of quadruples entering $T_4^{(\text{CIPSI})}$ and $\Lambda_4^{(\text{CIPSI})}$
are again extracted from $|\Psi^{(\text{CIPSI})}\rangle$.
\item[4.]
Use the information obtained in Step 3 to determine correction $\delta(P;Q)$,
Eq. (\ref{eq:mmcorrection}), which describes the remaining correlations of interest that were
not captured by the CC($P$) calculations. If the goal is to converge the CCSDT energetics,
define the $Q$ space needed to calculate $\delta(P;Q)$ as the remaining triply
excited determinants that are not contained in $|\Psi^{(\text{CIPSI})}\rangle$.
If the target approach is CCSDTQ, define the $Q$ space as the triply
and quadruply excited determinants absent in $|\Psi^{(\text{CIPSI})}\rangle$.
Add the resulting correction $\delta(P;Q)$ to $E^{(P)}$
to obtain the CC($P$;$Q$) energy $E^{(P+Q)}$, Eq. (\ref{eq:ccpq_energy}).
\item[5.]
To check convergence, repeat Steps 1--4 for a larger value of $N_{\text{det(in)}}$. The CIPSI-driven
CC($P$;$Q$) calculations can be regarded as converged if the difference between consecutive
$E^{(P+Q)}$ energies falls below a user-defined threshold. In analogy to the semi-stochastic
CC($P$;$Q$) framework of Refs.\
\onlinecite{stochastic-ccpq-prl-2017,stochastic-ccpq-jcp-2021,stochastic-ccpq-molphys-2020},
one can also stop if the fraction(s) of higher--than--doubly excited determinants
contained in the final CIPSI state $|\Psi^{(\text{CIPSI})}\rangle$ is (are) sufficiently
large to produce the desired accuracy level.
\end{itemize}

In this initial exploratory study, we implemented the CIPSI-driven CC($P$;$Q$) approach that allows us to converge
the CCSDT energetics. We did this by modifying our standalone CC($P$;$Q$) codes, described in Refs.\
\onlinecite{stochastic-ccpq-prl-2017,stochastic-ccpq-jcp-2021,stochastic-ccpq-molphys-2020,%
jspp-chemphys2012,jspp-jcp2012,jspp-jctc2012,nbjspp-molphys2017}
and interfaced with the RHF and integral transformation routines available in GAMESS,\cite{gamess,gamess2020} 
such that they can use the lists of triply excited determinants extracted from the CIPSI wave functions
$|\Psi^{(\text{CIPSI})}\rangle$, generated with Quantum Package 2.0, to set up the relevant $P$ spaces (as
already explained, the corresponding $Q$ spaces are automatically defined as the remaining triples absent in
the $|\Psi^{(\text{CIPSI})}\rangle$ wave functions). By design, as the input parameter $N_{\text{det(in)}}$
used to terminate CIPSI runs increases, producing longer and longer CI expansions to represent
wave functions $|\Psi^{(\text{CIPSI})}\rangle$, the CC($P$;$Q$) energies $E^{(P+Q)}$ approach their CCSDT
parents. The underlying CC($P$) calculations converge the CCSDT energetics too, but, as
further elaborated on in Section \ref{sec3},
by ignoring the triples that were not captured
by CIPSI, they do it at a much slower rate. In examining the convergence of the CIPSI-driven CC($P$) and
CC($P$;$Q$) energies toward CCSDT,
we sampled the $N_{\text{det(in)}}$ values in a roughly semi-logarithmic manner, starting from $N_{\text{det(in)}} = 1$.
Since all of the calculations reported in this work adopted RHF determinants as reference functions,
the $|\Psi^{(\text{CIPSI})}\rangle$ state becomes the RHF determinant and the resulting
CC($P$) and CC($P$;$Q$) energies become identical to those obtained in the RHF-based CCSD and CR-CC(2,3)
calculations, respectively, when $N_{\text{det(in)}} = 1$. Thus,
in analogy to the QMC propagation time $\tau$ used in our semi-stochastic CC($P$)/EOMCC($P$) and CC($P$;$Q$)
studies,\cite{stochastic-ccpq-prl-2017,eomccp-jcp-2019,stochastic-ccpq-molphys-2020,stochastic-ccpq-jcp-2021}
we can regard the $N_{\text{det(in)}}$ input variable defining CIPSI computations as the parameter connecting
CCSD [in the CC($P$) case] or CR-CC(2,3) [in the case of CC($P$;$Q$) runs] with CCSDT.
As a result, similarly to CCSD, CR-CC(2,3), and CCSDT, the CIPSI-driven CC($P$) and CC($P$;$Q$)
approaches considered in this work remain size extensive for all values of $N_{\text{det(in)}}$.
The CC($P$) calculations are size extensive, since they are nothing else than the usual CC
computations in which we solve the connected amplitude equations, Eq. (\ref{eq:cceqns}), for
the cluster operator $T^{(P)}$ defined by Eq. (\ref{eq:clusterop_p}). In the case of the
CIPSI-driven CC($P$) method implemented in this study, $T^{(P)} = T_1 + T_2 + T_3^{(\text{CIPSI})}$, where
$T_3^{(\text{CIPSI})} = \sum_{|\Phi_{ijk}^{abc}\rangle \in |\Psi^{(\text{CIPSI})}\rangle} t_{abc}^{ijk} \: E_{ijk}^{abc}$
is the $T_{3}$ operator defined using the list of triply excited determinants $|\Phi_{ijk}^{abc}\rangle$ contained
in the final CIPSI state $|\Psi^{(\text{CIPSI})}\rangle$ (we use the usual notation in which
$i,j,k$ and $a,b,c$ designate the occupied and unoccupied spin-orbitals in $|\Phi\rangle$, respectively, and
$E_{ijk}^{abc}$ is the elementary triple excitation operator generating
$|\Phi_{ijk}^{abc}\rangle$ from $|\Phi\rangle$). The
noniterative correction $\delta(P;Q)$, Eq. (\ref{eq:mmcorrection}), which in the case of the CIPSI-driven CC($P$;$Q$)
approach developed in this work captures the $T_{3}$ effects not described by $T_3^{(\text{CIPSI})}$ and
which involves the summation over the remaining triply excited determinants that are
not included in $|\Psi^{(\text{CIPSI})}\rangle$, i.e.,
$\delta(P;Q) = \sum_{|\Phi_{ijk}^{abc}\rangle \notin |\Psi^{(\text{CIPSI})}\rangle}
\ell_{ijk}^{abc} \: \mathfrak{M}_{abc}^{ijk}$,
being the connected quantity similar to that used in the CR-CC(2,3) and CC(t;3) methods,
is size extensive too (for the early numerical illustration of the size extensivity of CR-CC(2,3), see
Ref. \onlinecite{crccl_cpl}).

The numerical demonstration of the size extensivity of the
CIPSI-driven CC($P$) and CC($P$;$Q$) methods implemented in this work is shown in Table \ref{table1}.
Our example is the noninteracting ${\rm F}_{2} + {\rm Ne}$ system, described by the cc-pVDZ basis set,\cite{ccpvnz}
obtained by placing the Ne atom at 1,000 bohr from the stretched fluorine molecule in which the F--F bond length
was set at twice its equilibrium value to increase the magnitude of $T_{3}$ correlations.
Along with the ${\rm F}_{2} + {\rm Ne}$ system, we consider the
isolated ${\rm F}_{2}$ molecule having the same geometry as in ${\rm F}_{2} + {\rm Ne}$ and the isolated neon atom,
both described by the cc-pVDZ basis. The CIPSI
diagonalization sequence used to provide the list of triply excited determinants for the inclusion in the
$P$ space corresponding to the ${\rm F}_{2} + {\rm Ne}$ system was initiated from the RHF reference
determinant and defined by setting the wave function
termination parameter $N_{\text{det(in)}}$, the input parameter $f$ controlling the CIPSI wave function growth,
and the MBPT-based stopping parameter $\eta$ at 5,000, 2, and $10^{-6}$ hartree, respectively.
Following the above description, the $P$ space used in the CIPSI-driven CC($P$) calculation for the noninteracting
${\rm F}_{2} + {\rm Ne}$ dimer consisted of all singly and doubly excited determinants and a subset
of triply excited determinants contained in the last $|\Psi^{(\text{CIPSI})}\rangle$ state of the
$N_{\text{det(in)}} = 5,000$ CIPSI run. The $Q$ space needed to compute the corresponding CC($P$;$Q$)
correction $\delta(P;Q)$ was defined as the remaining triples not included in $|\Psi^{(\text{CIPSI})}\rangle$.
To ensure the consistency of the $P$ spaces used in the CC($P$) calculations for the ${\rm F}_{2} + {\rm Ne}$
system and the ${\rm F}_{2}$ and Ne fragments, we generated the $P$ space for ${\rm F}_{2}$ by
removing the triply excited determinants involving the orbitals of Ne
from the list of triples obtained in the $N_{\text{det(in)}}=5,000$ CIPSI run for ${\rm F}_{2} + {\rm Ne}$.
Similarly, the $P$ space used in the CC($P$) calculations for Ne was obtained by
starting from the list of triples produced in the $N_{\text{det(in)}}=5,000$ CIPSI calculation for
${\rm F}_{2} + {\rm Ne}$ and
removing the triply excited determinants involving the orbitals of ${\rm F}_{2}$.
As in all CC($P$;$Q$) calculations considered in this work, the $Q$ spaces associated with the ${\rm F}_{2}$ and Ne
monomers were defined as the remaining triples missing in the respective $P$ spaces. We chose the
$N_{\text{det(in)}} = 5,000$ value in the size extensivity test reported in Table \ref{table1}, since
it is sufficiently large to
introduce the leading triply excited determinants into the relevant $P$ spaces, while being small enough to
produce the CC($P$) and CC($P$;$Q$) energies that are visibly different than their CCSDT counterparts.

It is clear from the results presented in Table \ref{table1} that, in analogy to CCSD, CR-CC(2,3), and CCSDT,
the CIPSI-driven CC($P$) and CC($P$;$Q$) methods are size extensive.
Indeed, the CC($P$) and CC($P$;$Q$) energies of the ${\rm F}_{2} + {\rm Ne}$ dimer are numerically identical
to the corresponding sums of the energies of the ${\rm F}_{2}$ and Ne monomers. We observe the same behavior
for other values of the CIPSI wave function termination parameter $N_{\text{det(in)}}$. One may ask a question
why the interfragment triply excited determinants $|\Phi_{ijk}^{abc}\rangle$ having spin-orbital indices located on
different monomers, which are present in the final CIPSI state $|\Psi^{(\text{CIPSI})}\rangle$ of the noninteracting
${\rm F}_{2} + {\rm Ne}$ system and thus end up in the corresponding $P$ space, do not result in the violation
of size extensivity in the CC($P$) and CC($P$;$Q$) calculations. The answer to this question is that
the connected triply excited cluster amplitudes $t_{abc}^{ijk}$ carrying indices located on different
noninteracting fragments vanish when obtained by solving the explicitly connected CC($P$)
amplitude equations, Eq. (\ref{eq:cceqns}). We did not remove such interfragment $t_{abc}^{ijk}$ amplitudes
from our CC($P$) calculations for the ${\rm F}_{2} + {\rm Ne}$ system and confirmed that they do indeed vanish.
The use of CI diagonalizations in constructing the $P$ spaces for the CC($P$) and
CC($P$;$Q$) computations does not affect the size extensivity of the CIPSI-driven CC($P$) and CC($P$;$Q$)
approaches, since all we need from these diagonalizations are the lists of higher--than--doubly
excited determinants relevant to the CC theory of interest (in our case, where we target the CCSDT
energetics, the lists of triples), not the
CI excitation amplitudes themselves. For example,
as in all conventional CC calculations, the contributions from the interfragment triply excited determinants
$|\Phi_{ijk}^{abc}\rangle$ to the ground-state wave function of the noninteracting ${\rm F}_{2} + {\rm Ne}$
dimer are represented in the CC($P$) calculations by the disconnected $T_{1} T_{2}$ and $(1/6) T_{1}^{3}$ clusters.
The noniterative correction $\delta(P;Q)$, which provides information about those $T_{3}$ correlations
that were not captured by the preceding CC($P$) run, becomes the sum of the $\delta(P;Q)$ values
for the isolated ${\rm F}_{2}$ and Ne fragments.

Aside from size extensivity, as analyzed above, and high efficiency in converging the parent CCSDT energetics
discussed in Section \ref{sec3}, and in analogy to the active-orbital-based
\cite{jspp-chemphys2012,jspp-jcp2012,jspp-jctc2012,nbjspp-molphys2017}
and semi-stochastic
\cite{stochastic-ccpq-prl-2017,stochastic-ccpq-jcp-2021,stochastic-ccpq-molphys-2020}
CC($P$;$Q$) approaches, the CIPSI-driven CC($P$;$Q$) methodology examined in this work offers
significant savings in the computational effort compared to full CCSDT.
This is largely related to the fact that, as shown in Section \ref{sec3},
the convergence of the CIPSI-driven CC($P$;$Q$) energies toward their CCSDT parents
with the wave function termination parameter $N_{\text{det(in)}}$, with the number
of determinants used to generate the final CIPSI state $N_{\text{det(out)}}$, and
with the fractions of triples in the $P$ space captured by the CIPSI algorithm is very fast.
Indeed, the CPU times associated with the CIPSI runs using smaller
$N_{\text{det(in)}}$ values, resulting in smaller diagonalization spaces,
are relatively short compared to the converged CIPSI computations.
Next, as explained in detail in Refs.
\onlinecite{stochastic-ccpq-prl-2017,stochastic-ccpq-jcp-2021,stochastic-ccpq-molphys-2020},
the CC($P$) calculations using small fractions of triples in the $P$ space, which is
all one needs to converge the CCSDT-level energetics in the CIPSI-driven CC($P$;$Q$) runs, are
much faster than the corresponding CCSDT computations. Finally, as also explained in Refs.
\onlinecite{stochastic-ccpq-prl-2017,stochastic-ccpq-jcp-2021,stochastic-ccpq-molphys-2020},
the computational cost of determining the CC($P$;$Q$) correction $\delta(P;Q)$ is
less than the cost of a single iteration of CCSDT.

In examining the CIPSI-driven CC($P$) and CC($P$;$Q$) energies
in Section \ref{sec3},
we are primarily interested
in how fast they converge toward their parent CCSDT values as $N_{\text{det(in)}}$ and the fraction of
triples in the $P$ space increase. In the case of the $E_{\text{var}}$, $E_{\text{var}} + \Delta E^{(2)}$,
and $E_{\text{var}} + \Delta E_{\text{r}}^{(2)}$ energies, we do what is often done in CIPSI calculations
(see, e.g., Refs. \onlinecite{cipsi_2,cipsi_benzene}) and compare them to their counterparts obtained by
extrapolating the data obtained in the CIPSI runs defined by the largest $N_{\text{det(in)}}$
values to the FCI limit. Specifically, following the procedure used in Ref.\ \onlinecite{cipsi_benzene},
we performed a linear fit of the last four $E_{\text{var},k}+\Delta E_{\text{r},k}^{(2)}$ energies
leading to the final $|\Psi^{(\text{CIPSI})}\rangle$ state obtained for the largest value of
$N_{\text{det(in)}}$ in a given CIPSI sequence, plotted against the corresponding $\Delta E_{\text{r},k}^{(2)}$
corrections, and extrapolated the resulting line to the $\Delta E_{\text{r},k}^{(2)} = 0$ limit.


\section{Numerical Examples}
\label{sec3}

\begin{figure}[ht]
\includegraphics[width=\columnwidth]{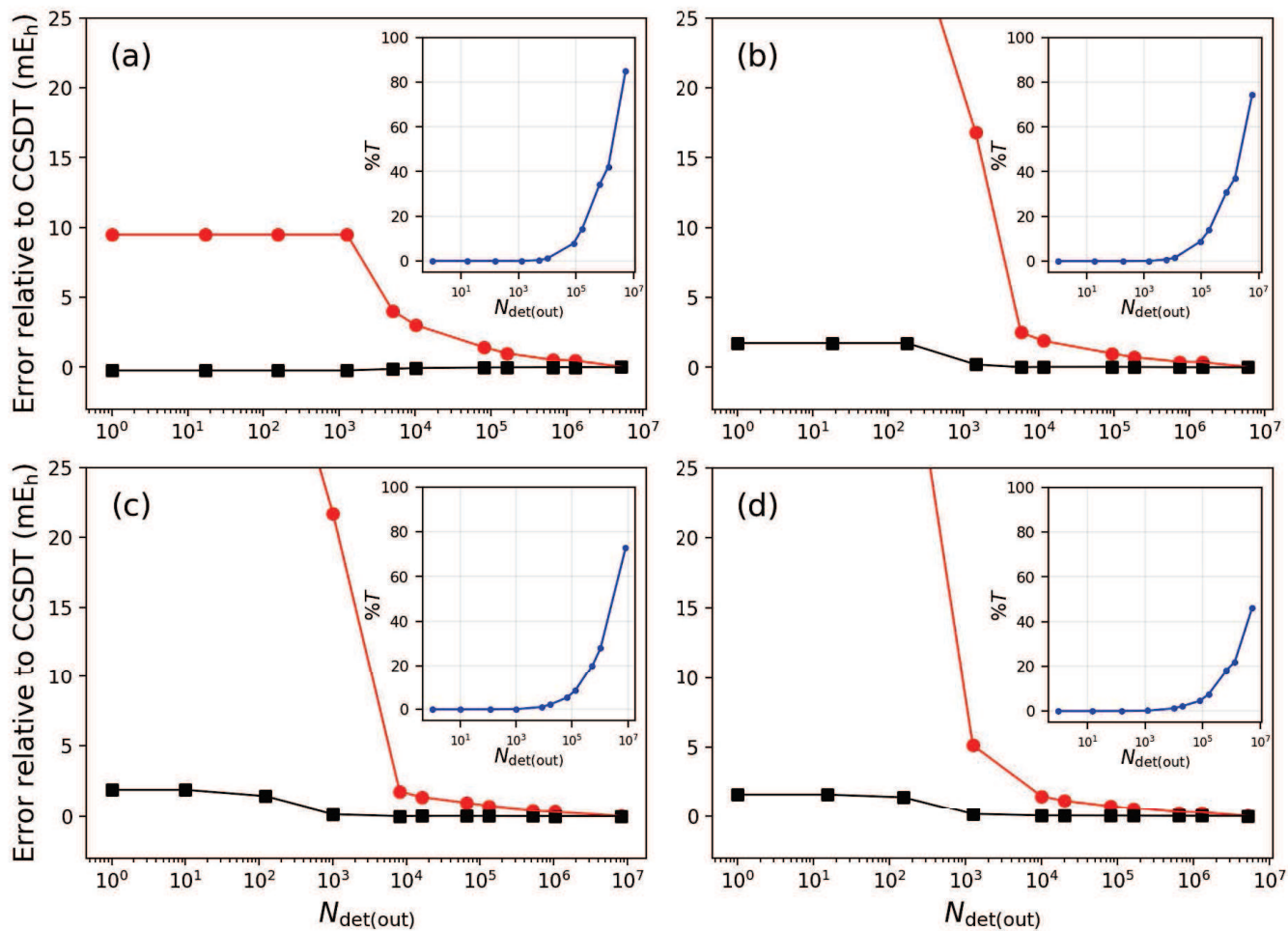}
\caption{Convergence of the CC($P$) (red lines and circles) and CC($P$;$Q$) (black lines and squares) 
energies toward their CCSDT parents as functions of the actual numbers of determinants,
$N_\mathrm{det(out)}$, defining the sizes of the final wave functions $|\Psi^{(\text{CIPSI})}\rangle$
generated in the underlying CIPSI runs, for the $\mathrm{F_2}$/cc-pVDZ molecule in which the F--F bond
length $R$ was set at (a) $R_\mathrm{e}$, (b) 1.5$R_\mathrm{e}$, (c) 2$R_\mathrm{e}$, and (d) 5$R_\mathrm{e}$,
where $R_e = 2.66816$ bohr is the equilibrium geometry.
The $P$ spaces used in the CC($P$) calculations consisted of all singles and doubles
and subsets of triples contained in the final $|\Psi^{(\text{CIPSI})}\rangle$ states of
the underlying CIPSI runs, whereas the $Q$ spaces needed to compute the corresponding CC($P$;$Q$)
corrections $\delta(P;Q)$ were defined as the remaining triples absent
in $|\Psi^{(\text{CIPSI})}\rangle$. The insets show the percentages of triples captured by the
CIPSI runs as functions of $N_\text{det(out)}$.}
\label{fig:figure1}
\end{figure}

We illustrate potential benefits offered by the CIPSI-driven CC($P$;$Q$) methodology,
when applied to recovering the CCSDT energetics, using a few molecular examples.
Our first example is the frequently studied dissociation of the fluorine molecule, as described by
the cc-pVDZ basis set.
We chose this example, since it is well established that
the CCSDT approach provides an accurate description of bond breaking in ${\rm F}_{2}$ (cf., e.g.,
Refs.\ \onlinecite{crccl_jcp,crccl_cpl,f2bh,musial-bartlett-jcp-2005}) and since we previously used it
to benchmark the CC($P$;$Q$)-based CC(t;3) approach\cite{jspp-chemphys2012} and the semi-stochastic
CC($P$;$Q$) methods driven by CIQMC\cite{stochastic-ccpq-prl-2017,stochastic-ccpq-jcp-2021}
and CCMC\cite{stochastic-ccpq-prl-2017} propagations. The results of our calculations for
the $\mathrm{F_2}$/cc-pVDZ system, in which the F--F bond length $R$ was stretched from its
equilibrium, $R_e = 2.66816$ bohr, value, where electron correlation effects are largely dynamical
in nature, to $1.5 R_e$, $2 R_e$, and $5 R_e$, where they gain an increasingly nondynamical
character, are summarized in Table \ref{tab:table2} and Fig.\ \ref{fig:figure1}. The complexity
of electron correlations in $\mathrm{F_2}$ manifests itself in the rapidly growing magnitude
of $T_{3}$ contributions as the F--F distance increases, as exemplified by the unsigned differences
between the CCSDT and CCSD energies, which are 9.485 millihartree at $R = R_{e}$,
32.424 millihartree at $R = 1.5 R_{e}$, 45.638 millihartree at $R =2 R_{e}$, and 49.816 millihartree
at $R = 5R_{e}$, when the cc-pVDZ basis set is employed. They grow with $R$ so fast that in the
$R = 2R_{e} - 5R_{e}$ region they become larger than the depth of the CCSDT potential (estimated at
$\sim$44 millihartree when the CCSDT energy at $R = R_{e}$ is subtracted from its $R = 5 R_{e}$
counterpart) and highly nonperturbative. The latter feature of $T_{3}$ contributions in the
stretched $\mathrm{F_2}$ molecule can be seen by examining the errors
relative to CCSDT obtained in the CCSD(T) calculations at $R = 1.5 R_{e}$, $2R_{e}$, and $5R_{e}$,
which are $-5.711$, $-23.596$, and $-39.348$ millihartree, respectively, when the cc-pVDZ basis set
is used. As shown in Table \ref{tab:table2}
[see the $N_{\text{det(in)}} = 1$ CC($P$;$Q$) energies], the CR-CC(2,3) triples correction to CCSD helps,
reducing the large errors characterizing CCSD(T)
to 1.735 millihartree at $R = 1.5 R_{e}$, 1.862 millihartree at $R =2 R_{e}$,
and 1.613 millihartree at $R = 5R_{e}$, which are much more acceptable, but, as demonstrated in our
earlier active-orbital-based and semi-stochastic CC($P$;$Q$) studies,
\cite{jspp-chemphys2012,stochastic-ccpq-prl-2017,stochastic-ccpq-jcp-2021} further error reduction
requires the relaxation of $T_{1}$ and $T_{2}$ clusters in the presence of the dominant $T_{3}$ contributions.
This is what the CIPSI-driven CC($P$;$Q$) methodology, where we use CIPSI runs
to identify the leading triple excitations for the inclusion in the $P$ space,
allows us to do. 

Indeed, with as little as 1,006--1,442 $S_{z} = 0$ determinants of the $A_{g}(D_{2h})$ symmetry
in the final Hamiltonian diagonalization spaces (we used $D_{2h}$ group, which is the largest Abelian
subgroup of the $D_{\infty h}$ symmetry group of $\mathrm{F_2}$, in our calculations),
generated by the inexpensive $N_{\text{det(in)}} = 1,000$ CIPSI runs at $R = 1.5 R_{e}$, $2R_{e}$, and $5R_{e}$,
which capture very small fractions, on the order of 0.1--0.2 \%, of all triples, the errors in the
resulting CC($P$;$Q$) energies relative to CCSDT are 0.202 millihartree at $R = 1.5 R_{e}$, 0.132 millihartree
at $R = 2 R_{e}$, and 0.144  millihartree at $R = 5 R_{e}$. This is an approximately tenfold error
reduction compared to the CR-CC(2,3) calculations, in which $T_{1}$ and $T_{2}$ clusters, obtained
with CCSD, are decoupled from $T_{3}$, and an improvement of the faulty CCSD(T) energetics by orders
of magnitude. As explained in detail in our papers on the CIQMC/CCMC-driven CC($P$;$Q$) approaches,
\cite{stochastic-ccpq-prl-2017,stochastic-ccpq-jcp-2021,stochastic-ccpq-molphys-2020}
with the fractions of triples in the relevant $P$ spaces being so small, the post-CIPSI steps of the
CC($P$;$Q$) calculations are not much more expensive than the CCSD-based CR-CC(2,3) computations
and a lot faster than the corresponding CCSDT computations.
The CC($P$;$Q$) calculations using $N_{\text{det(in)}} = 1,000$ do not offer any improvements over
CR-CC(2,3) at the equilibrium geometry, since the final diagonalization space of the underlying CIPSI run
does not yet contain any triply excited determinants, and the CR-CC(2,3) energy at $R = R_{e}$ is already
very accurate anyway, but with the relatively small additional effort corresponding to $N_{\text{det(in)}} = 10,000$,
which results in 10,150 $S_{z} = 0$ determinants of the $A_{g}(D_{2h})$ symmetry in the final CIPSI
diagonalization space and only 1.2 \% of all triples in the $P$ space, the unsigned error in the CC($P$;$Q$)
energy relative to its CCSDT parent substantially decreases, from 0.240 millihartree, when
$N_{\text{det(in)}} \leq 1,000$, to 67 microhartree, when $N_{\text{det(in)}}$ is set at 10,000.
The use of $N_{\text{det(in)}} = 10,000$ for
the remaining three geometries considered in Table \ref{tab:table2} and Fig.\ \ref{fig:figure1} produces
similarly compact $|\Psi^{(\text{CIPSI})}\rangle$ wave functions, spanned by 11,578--19,957 determinants,
similarly small fractions of triples in the corresponding $P$ spaces, ranging from 1.5 \% at $R = 1.5 R_{e}$
to 2.2 \% at  $R = 5 R_{e}$, and even smaller errors in the CC($P$;$Q$) energies relative to CCSDT.

It is clear from Table \ref{tab:table2} and Fig.\ \ref{fig:figure1} that the convergence of the
CIPSI-driven CC($P$;$Q$) energies toward CCSDT with the wave function termination parameter
$N_{\text{det(in)}}$, with the number of determinants used to
generate the final CIPSI state $|\Psi^{(\text{CIPSI})}\rangle$ [$N_{\text{det(out)}}$], and
with the fraction of triples in the $P$ space captured by the CIPSI procedure is very fast. The
uncorrected CC($P$) energies converge to CCSDT too, but they do it at a considerably slower rate
than their CC($P$;$Q$) counterparts. For example, the CIPSI-driven CC($P$) calculations reduce the
9.485, 32.424, 45.638, and 49.816 millihartree errors relative to CCSDT obtained with CCSD to 1.419,
0.991, 0.922, and 0.764 millihartree, respectively, when $N_{\text{det(in)}} = 50,000$, which translates
in the $N_{\text{det(out)}}$ values ranging between 65,172 and 92,682 and about 5--9 \% of all triples
included in the underlying $P$ spaces, but the errors characterizing
the corresponding CC($P$;$Q$) energies are already at the level of 20--30 microhartree at this stage,
which is obviously a substantial improvement over the CC($P$) results. It is also worth noticing that the
convergence of the CIPSI-driven CC($P$) and CC($P$;$Q$) energies toward their CCSDT parents with
$N_{\text{det(in)}}$ [or $N_{\text{det(out)}}$] is considerably faster than the convergence of the corresponding
variational and perturbatively corrected CIPSI energies toward the extrapolated
$E_{\text{var}} + \Delta E_{\text{r}}^{(2)}$ values. This is in line with the above observations
that indicate that the CIPSI-driven CC($P$;$Q$) calculations are capable of recovering the parent
CCSDT energetics, even when electronic quasi-degeneracies and $T_{3}$ clusters become significant,
out of the unconverged CIPSI runs that rely on relatively small diagonalization spaces. We observed
similar patterns when comparing the semi-stochastic, CIQMC- and CCMC-driven, CC($P$)/EOMCC($P$)
and CC($P$;$Q$) calculations with the underlying CIQMC/CCMC propagations.
\cite{stochastic-ccpq-prl-2017,eomccp-jcp-2019,stochastic-ccpq-molphys-2020,stochastic-ccpq-jcp-2021}

In analogy to the previously considered deterministic, active-orbital-based
\cite{jspp-chemphys2012,jspp-jcp2012,nbjspp-molphys2017,ccpq-be2-jpca-2018,ccpq-mg2-mp-2019}
and semi-stochastic, CIQMC/CCMC-based
\cite{stochastic-ccpq-prl-2017,stochastic-ccpq-jcp-2021} CC($P$;$Q$) studies,
the convergence of the CIPSI-driven CC($P$;$Q$) computations toward the parent
CCSDT energetics remains equally rapid when we use basis sets larger than cc-pVDZ.
This is illustrated in Table \ref{tab:table3}, where we show the
results of the CIPSI-driven CC($P$;$Q$) calculations for the $\mathrm{F_2}$
molecule at $R = 2 R_{e}$ using the cc-pVTZ basis set.\cite{ccpvnz}
As pointed out in Ref.\ \onlinecite{stochastic-ccpq-jcp-2021}, and
in analogy to the cc-pVDZ basis, the $T_{3}$ contribution characterizing the stretched
$\mathrm{F_2}$/cc-pVTZ system in which the internuclear separation is set at twice
the equilibrium bond length, estimated by forming the difference between the CCSDT and CCSD energies
at $-62.819$ millihartree, is not only very large, but also larger, in absolute value, than the
corresponding CCSDT dissociation energy, which is about 57 millihartree, when
the CCSDT energy at $R_{e}$ is subtracted from its $5R_{e}$ counterpart.
It is also highly nonperturbative at the same time, as
demonstrated by the $-26.354$ millihartree error relative to CCSDT obtained with CCSD(T).
Again, the CR-CC(2,3) triples correction to CCSD, equivalent to the $N_{\text{det(in)}} = 1$
CC($P$;$Q$) calculation in Table \ref{tab:table3}, works a lot better than CCSD(T),
but the 4.254 millihartree error relative to CCSDT remains. With as little as 5,118
$S_{z} = 0$ determinants of the $A_{g}(D_{2h})$ symmetry in the final diagonalization space
obtained by the nearly effortless $N_{\text{det(in)}} = 5,000$ CIPSI run, which captures
0.03 \% of all triples, the difference between the CC($P$;$Q$) and CCSDT energies decreases
to 0.345 millihartree, and with the help of the $N_{\text{det(in)}} = 50,000$ CIPSI calculation,
which is still relatively inexpensive, resulting in 82,001 $S_{z} = 0$ $A_{g}(D_{2h})$-symmetric
determinants in the final diagonalization space and less than 1 \% of the triples in the $P$ space,
the error in the CC($P$;$Q$) energy relative to its CCSDT parent reduces to less than 0.1 millihartree.
Similarly to the cc-pVDZ basis, the convergence of the CIPSI-driven CC($P$;$Q$)
energies toward CCSDT with $N_{\text{det(in)}}$, $N_{\text{det(out)}}$, and the fraction of triples
in the $P$ space captured by the CIPSI algorithm is not only fast, when the larger cc-pVTZ basis set
is employed, but also much faster than in the case of the uncorrected CC($P$) calculations.
Once again, as $N_{\text{det(in)}}$ increases, the rate of convergence of the CIPSI-driven
CC($P$) and CC($P$;$Q$) energies toward their CCSDT parent is higher than those
characterizing the corresponding variational and perturbatively corrected CIPSI energies
in their attempt to recover the extrapolated $E_{\text{var}} + \Delta E_{\text{r}}^{(2)}$ energy.

Our final test, shown in Table \ref{tab:table4}, is the frequently examined
\cite{jspp-jcp2012,balkova1994,tailored3,CBDexp,carpenter1983,hess1983,goddard1986,carsky1988,%
demel-pittner-nonit,MR-AQCC,mkcc_our1,karadakov2008,icmrcc6,BCCC5,paldus-li-succ-2009,%
stochastic-ccpq-prl-2017,evangelista-mrsrg-2019,hfs-aci-ec-cc-2021,stochastic-ccpq-jcp-2021}
automerization of cyclobutadiene. In this case, one of the key challenges is an accurate
determination of the activation barrier, which requires a well-balanced description of the
nondegenerate, rectangle-shaped, closed-shell reactant (or the equivalent product) species, in which electron
correlation effects are largely dynamical in nature, and the quasi-degenerate, square-shaped, transition state
characterized by substantial nondynamical correlations associated with its strongly diradical character.
Experimental estimates of the activation barrier for the automerization of cyclobutadiene, which
range from 1.6 to 10 kcal/mol,\cite{CBDexp,carpenter1983} are not very precise, but the most accurate
single- and multi-reference {\it ab initio} computations, compiled, for example, in Refs.\
\onlinecite{jspp-jcp2012,tailored3,evangelista-mrsrg-2019}, place it in the 6--10 kcal/mol range.
This, in particular, applies to the CCSDT approach,\cite{balkova1994,jspp-jcp2012} which is of the
primary interest in the present study. Indeed, if we, for example, use the reactant
and transition-state geometries obtained with the multi-reference average-quadratic CC
(MR-AQCC) approach\cite{mraqcc1,mraqcc2} in Ref.\ \onlinecite{MR-AQCC} and the cc-pVDZ basis set,
the CCSDT value of the activation energy characterizing the automerization of cyclobutadiene becomes
7.627 kcal/mol,\cite{jspp-jcp2012} in good agreement with the best {\it ab initio} calculations carried out to
date. By adopting the same geometries and basis set in this initial benchmark study of the CIPSI-driven CC($P$;$Q$)
methodology, we can examine if the CC($P$;$Q$) calculations using the $P$ spaces constructed with the
help of CIPSI runs are capable of converging this result. The main challenge here is that the $T_{3}$
effects, estimated as the difference between the CCSDT and CCSD energies, are not only very
large, but also hard to balance. When the cc-pVDZ basis set used in this study is
employed, they are $-26.827$ millihartree for the reactant and $-47.979$ millihartree for the transition state.
Furthermore, in the case of the transition state, the coupling of the lower-rank $T_{1}$ and $T_{2}$ clusters
with their higher-rank $T_{3}$ counterpart is so large that none of the noniterative triples corrections
to CCSD provide a reasonable description of the activation barrier.\cite{jspp-jcp2012,balkova1994,tailored3}
This, in particular, applies to the CR-CC(2,3) approach, equivalent to the $N_{\text{det(in)}} = 1$
CC($P$;$Q$) calculation, which produces an activation barrier exceeding 16 kcal/mol, when the cc-pVDZ basis set
is employed, instead of less than 8 kcal/mol obtained with CCSDT (cf. Table \ref{tab:table4}). The failure
of CR-CC(2,3) to provide an accurate description of the activation energy is a consequence of its
inability to accurately describe the transition state. Indeed, the difference between the CR-CC(2,3)
and CCSDT energies at the transition-state geometry is 14.636 millihartree, when the cc-pVDZ basis set is
employed, as opposed to only 0.848 millihartree obtained for the reactant. As discussed in detail in
Refs.\ \onlinecite{jspp-jcp2012,tailored3}, other triples corrections to CCSD, including the widely used
CCSD(T) approach, face similar problems. We demonstrated in Refs.\
\onlinecite{jspp-jcp2012,stochastic-ccpq-prl-2017,stochastic-ccpq-jcp-2021} that the deterministic
CC($P$;$Q$)-based CC(t;3) approach and the semi-stochastic CC($P$;$Q$) calculations using CIQMC and CCMC
are capable of accurately approximating the CCSDT energies of the reactant and transition-state species
and the CCSDT activation barrier, so it is interesting to explore if the CIPSI-driven CC($P$;$Q$)
methodology can do the same.

As shown in Table \ref{tab:table4}, the CC($P$;$Q$) calculations using CIPSI to identify the
dominant triply excited determinants for the inclusion in the $P$ space are very efficient
in converging the CCSDT energetics. With the final diagonalization spaces spanned by a little over
110,000 $S_{z} = 0$ determinants of the $A_{g}(D_{2h})$ symmetry (we used $D_{2h}$ group
for both the $D_{2h}$-symmetric reactant and the $D_{4h}$-symmetric transition state in our calculations),
generated in the relatively inexpensive CIPSI runs defined by $N_{\text{det(in)}} = 100,000$ that
capture 0.1 \% of all triples, the 0.848 millihartree, 14.636 millihartree, and 8.653 kcal/mol
errors in the reactant, transition-state, and activation energies relative to CCSDT obtained with CR-CC(2,3)
are reduced by factors of 2--4, to 0.382 millihartree, 3.507 millihartree, and 1.961 kcal/mol, respectively,
when the CC($P$;$Q$) approach is employed. When $N_{\text{det(in)}}$ is increased to 500,000,
resulting in about 890,000--900,000 $S_{z} = 0$ determinants of the $A_{g}(D_{2h})$ symmetry
in the final diagonalization spaces used by CIPSI and 1.0--1.2 \% of the triples in the resulting $P$ spaces,
the errors in the CC($P$;$Q$) reactant, transition-state, and activation energies relative to
CCSDT become 0.267 millihartree, 0.432 millihartree, and 0.104 kcal/mol. Clearly, these are
great improvements compared to the initial $N_{\text{det(in)}} = 1$, i.e., CR-CC(2,3),
values, especially if we realize that with the fractions of triples being so small,
the post-CIPSI steps of the CC($P$;$Q$) computations are not only a lot faster than the parent CCSDT runs,
but also not much more expensive than the corresponding CR-CC(2,3) calculations, as elaborated on in Refs.\
\onlinecite{stochastic-ccpq-prl-2017,stochastic-ccpq-jcp-2021,stochastic-ccpq-molphys-2020}.

In analogy to the previously discussed case of bond breaking in $\mathrm{F_2}$, the
convergence of the CIPSI-driven CC($P$;$Q$) energies toward CCSDT for the reactant and
transition-state species defining the automerization of cyclobutadiene with
$N_{\text{det(in)}}$, $N_{\text{det(out)}}$, and the fractions of triples in the relevant $P$ spaces
captured by the underlying CIPSI runs is not only very fast, but also significantly faster
than that characterizing the uncorrected CC($P$) calculations. For each of the two species,
the CC($P$) energies converge toward their CCSDT parent in a steady fashion, but, as shown in
Table \ref{tab:table4}, their convergence is rather slow, emphasizing the importance of
correcting the results of the CC($P$) calculations for the missing triple excitations
not captured by the CIPSI runs using smaller diagonalization spaces. Similarly to the
previously examined active-orbital-based
\cite{jspp-chemphys2012,jspp-jcp2012,nbjspp-molphys2017,ccpq-be2-jpca-2018,ccpq-mg2-mp-2019}
and CIQMC/CCMC-based
\cite{stochastic-ccpq-prl-2017,stochastic-ccpq-jcp-2021} CC($P$;$Q$) approaches,
our moment correction $\delta(P;Q)$, defined by Eq. (\ref{eq:mmcorrection}),
is very effective in this regard. Another interesting observation, which can be made
based on the results presented in Table \ref{tab:table4}, is that while the CC($P$) energies
for the individual reactant and transition-state species converge toward their CCSDT parent values in a
steady fashion, the corresponding activation barrier values behave in a less systematic manner,
oscillating between about $-1$ and 1 kcal/mol when $N_{\text{det(in)}} = 500,000\mbox{--}15,000,000$.
One might try to eliminate this behavior, which is a consequence of a different character
of the many-electron correlation effects in the reactant and transition-state species, by merging the
$P$ spaces used to perform the CC($P$) calculations for the two structures, but, as shown in
Table \ref{tab:table4}, the CC($P$;$Q$) correction $\delta(P;Q)$, which is highly effective
in capturing the missing $T_{3}$ correlations, takes care of this problem too. As
$N_{\text{det(in)}}$, $N_{\text{det(out)}}$, and the fractions of triples in the $P$ spaces
used in the CC($P$) calculations for the reactant and transition state increase, the
CC($P$;$Q$) values of the activation barrier converge toward its CCSDT parent rapidly and in
a smooth manner, eliminating, at least to a large extent, the need to equalize the $P$ spaces used in the
underlying CC($P$) steps. As in the case of bond breaking in the fluorine molecule,
the convergence of the CIPSI-driven CC($P$) and CC($P$;$Q$) energies toward
their CCSDT parents with $N_{\text{det(in)}}$/$N_{\text{det(out)}}$
is considerably faster than the convergence of the variational and
perturbatively corrected CIPSI energies toward the extrapolated
$E_{\text{var}} + \Delta E_{\text{r}}^{(2)}$ values, but we must keep in mind that
the calculated CCSDT and extrapolated $E_{\text{var}} + \Delta E_{\text{r}}^{(2)}$
energies, while representing the respective parent limits for the CC($P$;$Q$) and CIPSI
calculations, are fundamentally different quantities, especially when higher--than--triply
excited cluster components, which are not considered in this work, become significant.
As one might anticipate, the $N_{\text{det(in)}}$ values needed to accurately represent
the CCSDT energies of the reactant and transition-state species of cyclobutadiene
by the CIPSI-driven CC($P$;$Q$) approach
are considerably larger than those used in the analogous CC($P$;$Q$) calculations
for the smaller ${\rm F}_{2}$ system, but they are orders of magnitude smaller
than the values of $N_{\text{det(in)}}$ required to obtain the similarly well
converged $E_{\text{var}} + \Delta E_{\text{r}}^{(2)}$ energetics in the underlying
CIPSI runs.


\section{Conclusions}
\label{sec4}

Inspired by our recent studies\cite{stochastic-ccpq-prl-2017,stochastic-ccpq-molphys-2020,stochastic-ccpq-jcp-2021}
aimed at determining accurate electronic energies equivalent to the results of high-level CC calculations,
in which we combined the deterministic CC($P$;$Q$) framework
\cite{jspp-chemphys2012,jspp-jcp2012,jspp-jctc2012,nbjspp-molphys2017,ccpq-be2-jpca-2018} with the stochastic
CIQMC\cite{Booth2009,Cleland2010,fciqmc-uga-2019,ghanem_alavi_fciqmc_jcp_2019,ghanem_alavi_fciqmc_2020}
and CCMC\cite{Thom2010,Franklin2016,Spencer2016,Scott2017} propagations, and the successes of
modern formulations\cite{adaptive_ci_1,adaptive_ci_2,asci_1,asci_2,ici_1,ici_2,shci_1,shci_2,shci_3,
cipsi_1,cipsi_2} of selected CI techniques,\cite{sci_1,sci_2,sci_3,sci_4} we have proposed a new form
of the CC($P$;$Q$) approach, in which we identify the dominant higher--than--doubly excited determinants
for the inclusion in the underlying $P$ spaces with the help of the selected CI algorithm abbreviated as
CIPSI.\cite{sci_3,cipsi_1,cipsi_2}
To illustrate potential benefits offered by the proposed merger of the CC($P$;$Q$) and CIPSI methodologies,
we have implemented the CIPSI-driven CC($P$;$Q$) method designed to converge CCSDT energetics.
The advantages of the CIPSI-driven CC($P$;$Q$) methodology have been illustrated by a few
numerical examples, including bond breaking in ${\rm F}_{2}$ and the automerization
of cyclobutadiene, which are accurately described by CCSDT.

The reported benchmark calculations demonstrate that the convergence of the CIPSI-driven CC($P$;$Q$) energies
toward CCSDT with the wave function termination parameter $N_{\text{det(in)}}$ adopted by CIPSI, with the number
of determinants used to generate the final CIPSI state [$N_{\text{det(out)}}$],
and with the fractions of triples in the $P$ space captured by the CIPSI procedure is very fast. As a result,
one can obtain CCSDT-level energetics, even when electronic quasi-degeneracies and $T_{3}$ clusters become
substantial, based on the information extracted from the relatively inexpensive CIPSI runs. This can be
attributed to two key factors. The first one is a tempered wave function growth through iterative Hamiltonian
diagonalizations in the modern implementation of CIPSI available in Quantum Package 2.0,\cite{cipsi_1,cipsi_2}
which we utilized in this work, resulting in an economical selection of the dominant triply excited determinants
(in general, the dominant higher--than--doubly excited determinants) for the inclusion in the $P$ spaces
driving the CC($P$;$Q$) computations. The second one is the efficiency of the moment corrections $\delta(P;Q)$
defining the CC($P$;$Q$) formalism, which provide an accurate and robust description of the missing $T_{3}$
contributions that cannot be captured by the underlying CC($P$) calculations using small fractions of triples
identified by the CIPSI runs employing smaller diagonalization spaces.
We have also shown that the uncorrected CC($P$) energies converge
with $N_{\text{det(in)}}$, $N_{\text{det(out)}}$, and the fractions of triples in the $P$ spaces constructed
with the help of CIPSI to their CCSDT parent values too, but they do it at a much slower rate, so that
we do not recommend the uncorrected CC($P$) approach in calculations aimed at recovering
high-level CC energetics.

Clearly, the present study is only our initial exploration of the CIPSI-driven (or, in general, selected-CI-driven)
CC($P$;$Q$) methodology, which needs more work. In addition to code optimization and more numerical tests,
especially including larger molecules and basis sets, we would like to extend the proposed CIPSI-driven
CC($P$;$Q$) framework to higher levels of the CC theory, beyond CCSDT, as we already did in the context of
the active-orbital-based\cite{nbjspp-molphys2017,ccpq-be2-jpca-2018} and CIQMC-based\cite{stochastic-ccpq-jcp-2021}
CC($P$;$Q$) considerations, and examine if other selected CI methods, such as heat-bath CI\cite{shci_1,shci_2,shci_3}
or adaptive-CI,\cite{adaptive_ci_1, adaptive_ci_2} to mention a couple of examples, are as useful in the
context of CC($P$;$Q$) considerations as the CIPSI approach adopted in this work.
Following our recent semi-stochastic EOMCC($P$) and CC($P$;$Q$) work,
\cite{eomccp-jcp-2019,stochastic-ccpq-molphys-2020} we are also planning to extend the CIPSI-driven
CC($P$;$Q$) methodology to excited electronic states. One of the main advantages of CIPSI and other
selected-CI methods, which are based on Hamiltonian diagonalizations, is that they can be easily
applied to excited states (see, e.g., Refs.\
\onlinecite{adaptive_ci_2,shci_excbench,cipsi_excbench_1,cipsi_excbench_2,cipsi_excbench_3,%
cipsi_excbench_4,cipsi_excbench_5}). This would allow us to construct the state-specific $P$ spaces, adjusted
to the individual electronic states of interest, which is more difficult to accomplish within the
CIQMC framework (see Refs.\ \onlinecite{eomccp-jcp-2019,stochastic-ccpq-molphys-2020} for additional comments).
Encouraged by our recent work on the semi-stochastic CC($P$;$Q$) models using truncated CIQMC rather than
FCIQMC propagations to determine the underlying $P$ spaces,\cite{stochastic-ccpq-jcp-2021}
we would like to examine if one can replace
the unconstrained CIPSI algorithm used in this study, which explores the entire many-electron Hilbert space in the
iterative wave function build-up, by its less expensive
truncated analogs compatible with the determinantal spaces needed
in the CC calculations of interest (e.g., the CISDT or CISDTQ analogs of CIPSI if one is interested in
converging the CCSDT or CCSDTQ energetics through CIPSI-driven CC($P$;$Q$) computations).
This may help us to achieve the desired high accuracy levels in the CIPSI-driven CC($P$;$Q$)
calculations with the relatively short CI wave function expansions, even when larger systems are examined,
since the diagonalization spaces generated by the truncated CIPSI models will be significantly smaller
than those produced when CIPSI is allowed to explore the entire many-electron Hilbert space.
Last, but not least,
inspired by our recent work on the CIPSI-driven externally corrected CC models,\cite{im-kg-pp-jed-js-jctc-2021}
we would like to investigate the effect of the CIPSI input parameter $f$ that controls the wave function growth
in successive Hamiltonian diagonalizations, which was set in this study at the default value of 2, on the
rate of convergence of the CIPSI-driven CC($P$;$Q$) energies toward their high-level CC parents, such as those
obtained with CCSDT.

\begin{acknowledgments}
This work has been supported by the Chemical Sciences, Geosciences
and Biosciences Division, Office of Basic Energy Sciences, Office
of Science, U.S. Department of Energy (Grant No. DE-FG02-01ER15228 to P.P), and Phase
I and II Software Fellowships awarded to J.E.D. by the Molecular Sciences Software Institute
funded by the National Science Foundation grant ACI-1547580.
We thank Drs. Pierre-Fran{\c c}ois Loos and Anthony Scemama for useful
discussions about Quantum Package 2.0 employed in our CIPSI computations.
\end{acknowledgments}

\section*{Data Availability}

The data that support the findings of this study are available within the article.


\section*{References}
\bibliographystyle{aipnum4-1} 
\renewcommand{\baselinestretch}{1.1}
%

\newpage

\clearpage


\onecolumngrid


\squeezetable
\begin{table*}[h!]
\caption{\label{table1}
Numerical demonstration of the size extensivity of the CIPSI-driven CC($P$) and CC($P$;$Q$) approaches,
alongside the analogous CCSD, CR-CC(2,3), and CCSDT calculations,
using the noninteracting ${\rm F}_{2} + {\rm Ne}$ system, as described by the cc-pVDZ basis set,
in which the F--F bond length $R$ was fixed at twice its equilibrium value.
In all post-RHF calculations, the core orbitals correlating with the $1s$ shells of
the fluorine and neon atoms were frozen and the Cartesian components of $d$ orbitals were employed throughout.
All energy values are total electronic energies in hartree.
}
\footnotesize
\begin{ruledtabular}
\begin{tabular}{l c c c c }
        Method &
        $E(\mathrm{F_2}+\mathrm{Ne})$\footnotemark[1] &
        $E(\mathrm{F_2})$\footnotemark[2] &
        $E(\mathrm{Ne})$ &
        $E(\mathrm{F_2}+\mathrm{Ne}) - [E(\mathrm{F_2}) + E(\mathrm{Ne})]$ \\
        \colrule
CCSD\footnotemark[3] & $-327.692849962$ & $-199.012562571$ & $-128.680287394$ & 0.000000003 \\
CR-CC(2,3)\footnotemark[4] & $-327.737915219$ & $-199.056339293$ & $-128.681575920$ & $-0.000000006$ \\
CC($P$)/$N_{\text{det(in)}}=5,000$ & $-327.736961010$\footnotemark[5] & $-199.056233029$\footnotemark[6] & $-128.680728060$\footnotemark[7] & 0.000000080 \\
CC($P$;$Q$)/$N_{\text{det(in)}}=5,000$ & $-327.739651938$\footnotemark[5] & $-199.058190353$\footnotemark[6] & $-128.681461627$\footnotemark[7] & 0.000000040 \\
CCSDT & $-327.739605196$ & $-199.058201287$ & $-128.681403900$ & $-0.000000009$
\end{tabular}
\end{ruledtabular}

\footnotetext[1]{
\setlength{\baselineskip}{1em}
The noninteracting ${\rm F}_{2} + {\rm Ne}$ system was obtained by placing the Ne atom along the axis of the F--F bond
at 1,000 bohr from the center of mass of the stretched fluorine molecule in which the internuclear separation $R$
was set at $2 R_e$, where $R_e = 2.66816$ bohr is the equilibrium geometry of ${\rm F}_{2}$.
}
\footnotetext[2]{
\setlength{\baselineskip}{1em}
The stretched ${\rm F}_{2}$ molecule in which the F--F bond length $R$ was set at $2 R_e$.
}
\footnotetext[3]{
\setlength{\baselineskip}{1em}
Equivalent to the CC($P$) calculations with $N_{\text{det(in)}}=1$.
}
\footnotetext[4]{
\setlength{\baselineskip}{1em}
Equivalent to the CC($P$;$Q$) calculations with $N_{\text{det(in)}}=1$.
}
\footnotetext[5]{
\setlength{\baselineskip}{1em}
The $P$ space used in the CC($P$) calculation for the ${\rm F}_{2} + {\rm Ne}$ system consisted of all singles
and doubles and a subset of triples contained in the final $|\Psi^{(\text{CIPSI})}\rangle$ state of the
underlying $N_{\text{det(in)}} = 5,000$ CIPSI run. The $Q$ space needed to compute the CC($P$;$Q$) correction $\delta(P;Q)$
was defined as the remaining triples absent in $|\Psi^{(\text{CIPSI})}\rangle$.
The $N_{\text{det(in)}} = 5,000$ CIPSI run for ${\rm F}_{2} + {\rm Ne}$, which was initiated from the RHF reference determinant,
used $f = 2$ and $\eta = 10^{-6}$ hartree.
}
\footnotetext[6]{
\setlength{\baselineskip}{1em}
The $P$ space used in the CC($P$) calculation for $\mathrm{F_2}$ was obtained by
removing the triply excited determinants involving Ne orbitals from the list
of triples provided by the $N_{\text{det(in)}}=5,000$ CIPSI run for the
$\mathrm{F_2}+\mathrm{Ne}$ system. The $Q$ space needed to compute the corresponding CC($P$;$Q$)
correction $\delta(P;Q)$ was defined as the remaining triples missing in the $P$ space.
}
\footnotetext[7]{
\setlength{\baselineskip}{1em}
The $P$ space used in the CC($P$) calculation for Ne was obtained by
removing the triply excited determinants involving $\mathrm{F_2}$ orbitals from the list
of triples provided by the $N_{\text{det(in)}}=5,000$ CIPSI run for the
$\mathrm{F_2}+\mathrm{Ne}$ system. The $Q$ space needed to compute the corresponding CC($P$;$Q$) 
correction $\delta(P;Q)$ was defined as the remaining triples missing in the $P$ space.
}
\end{table*}


\squeezetable
\begin{table*}[h!]
\caption{\label{tab:table2}
Convergence of the CC($P$) and CC($P$;$Q$) energies toward CCSDT, alongside the
variational and perturbatively corrected CIPSI energies, for the
$\mathrm{F_2}$/cc-pVDZ molecule in which the F--F bond length $R$ was set at $R_e$, $1.5 R_e$, $2 R_e$, and $5 R_e$,
where $R_e = 2.66816$ bohr is the equilibrium geometry.
For each value of the wave function termination parameter $N_{\text{det(in)}}$,
the $P$ space used in the CC($P$) calculations consisted of all singles and doubles
and a subset of triples contained in the final $|\Psi^{(\text{CIPSI})}\rangle$ state of
the underlying CIPSI run, whereas the $Q$ space needed to compute the corresponding CC($P$;$Q$)
correction $\delta(P;Q)$ was defined as the remaining triples absent
in $|\Psi^{(\text{CIPSI})}\rangle$. In all post-RHF calculations, the two lowest-lying
core orbitals were frozen and the Cartesian components of $d$ orbitals were employed throughout.
Each CIPSI run was initiated from the RHF reference determinant and the MBPT-based stopping
parameter $\eta$ was set at $10^{-6}$ hartree. The input parameter $f$ controlling the
CIPSI wave function growth was set at the default value of 2.
}
\footnotesize
\begin{ruledtabular}
\begin{tabular}{l p{3.5cm} c c c c c c}
\textrm{$R/R_e$}&
\textrm{$N_{\text{det(in)}}$ / $N_{\text{det(out)}}$}&
\textrm{\% of triples}&
\textrm{$E_{\text{var}}$\footnotemark[1]}&
\textrm{$E_{\text{var}}+\Delta E^{(2)}$\footnotemark[1]}&
\textrm{$E_{\text{var}}+\Delta E_{\text{r}}^{(2)}$\footnotemark[1]}&
\textrm{CC$(P)$\footnotemark[2]}&
\textrm{CC$(P;Q)$\footnotemark[2]} \\
\colrule
	1.0 & 1 / 1 & 0 & 418.057\footnotemark[3] & $-94.150$\footnotemark[4] & $-12.651$ & 9.485\footnotemark[5] & $-0.240$\footnotemark[6] \\
	& 10 / 17               & 0    & 330.754 & $-32.707$   & -4.877    & 9.485 & $-0.240$ \\
	& 100 / 154             & 0    & 232.186 & $-7.963$    & 2.338     & 9.485 & $-0.240$ \\
	& 1,000 / 1,266         & 0    & 65.926  & 1.480     & 2.079     & 9.485 & $-0.240$ \\
	& 5,000 / 5,072         & 0.4  & 23.596  & $-0.133(0)$ & $-0.069(0)$ & 4.031 & $-0.129$ \\
	& 10,000 / 10,150       & 1.2  & 19.197  & 0.045(2)  & 0.084(2)  & 3.010 & $-0.067$ \\
	& 50,000 / 81,288       & 7.9  & 11.282  & 0.133(1)  & 0.145(1)  & 1.419 & $-0.031$ \\
	& 100,000 / 162,430     & 14.5 & 9.222   & 0.138(1)  & 0.146(1)  & 0.983 & $-0.020$ \\
	& 500,000 / 649,849     & 34.3 & 5.630   & 0.092(1)  & 0.095(1)  & 0.519 & $-0.009$ \\
	& 1,000,000 / 1,300,305 & 42.2 & 4.816   & 0.072(0)  & 0.074(0)  & 0.464 & $-0.008$ \\
	& 5,000,000 / 5,187,150 & 85.1 & 1.161   & 0.015(2)  & 0.016(2)  & 0.023 & $-0.001$ \\
        &                       &      &         &           &           &       &          \\ [-1.5mm]
	1.5 & 1 / 1 & 0 & 541.109\footnotemark[3] & $-130.718$\footnotemark[4] & 137.819 & 32.424\footnotemark[5] & 1.735\footnotemark[6] \\
	& 10 / 18               & 0    & 319.363 & $-11.279$  & 10.126   & 32.424 & 1.735 \\
	& 100 / 177             & 0    & 235.819 & 2.527    & 12.175   & 32.424 & 1.735 \\
	& 1,000 / 1,442         & 0.1  & 77.306  & 5.218    & 5.948    & 16.835 & 0.202 \\
	& 5,000 / 5,773         & 0.7  & 21.091  & 0.811(2) & 0.856(2) & 2.490  & 0.009 \\
	& 10,000 / 11,578       & 1.5  & 17.333  & 0.811(2) & 0.839(2) & 1.892  & 0.028 \\
	& 50,000 / 92,682       & 8.8  & 10.879  & 0.762(1) & 0.771(1) & 0.991  & 0.033 \\
	& 100,000 / 185,350     & 13.9 & 9.243   & 0.632(1) & 0.639(1) & 0.727  & 0.023 \\
	& 500,000 / 742,754     & 30.8 & 5.586   & 0.391(1) & 0.393(1) & 0.390  & 0.005 \\
	& 1,000,000 / 1,484,218 & 37.1 & 4.795   & 0.330(0) & 0.332(0) & 0.362  & 0.004 \\
	& 5,000,000 / 5,907,228 & 74.3 & 1.165   & 0.079(2) & 0.079(2) & 0.028  & $-0.000$ \\
        &                       &      &         &           &           &       &          \\ [-1.5mm]
	2.0 & 1 / 1 & 0 & 640.056\footnotemark[3] & $-159.482$\footnotemark[4] & 289.080 & 45.638\footnotemark[5] & 1.862\footnotemark[6] \\
	& 10 / 10               & 0    & 337.263 & $-3.392$   & 19.484   & 45.638 & 1.862 \\
	& 100 / 122             & 0.0  & 250.492 & 6.090    & 16.021   & 38.309 & 1.411 \\
	& 1,000 / 1,006         & 0.1  & 105.265 & 5.589    & 7.036    & 21.727 & 0.132 \\
	& 5,000 / 8,118         & 1.1  & 17.355  & 0.787(1) & 0.815(1) & 1.725  & $-0.003$ \\
	& 10,000 / 16,291       & 2.1  & 14.555  & 0.860(1) & 0.878(1) & 1.338  & 0.012 \\
	& 50,000 / 65,172       & 5.2  & 11.064  & 0.800(1) & 0.810(1) & 0.922  & 0.015 \\
	& 100,000 / 130,448     & 8.4  & 9.410   & 0.655(1) & 0.662(1) & 0.695  & 0.009 \\
	& 500,000 / 521,578     & 19.8 & 5.929   & 0.375(1) & 0.378(1) & 0.400  & 0.005 \\
	& 1,000,000 / 1,043,539 & 28.0 & 4.820   & 0.306(0) & 0.308(0) & 0.314  & 0.002 \\
	& 5,000,000 / 8,190,854 & 72.8 & 0.764   & 0.047(1) & 0.047(1) & 0.009  & $-0.000$ \\
        &                       &      &         &           &           &       &          \\ [-1.5mm]
	5.0 & 1 / 1 & 0 & 730.244\footnotemark[3] & $-183.276$\footnotemark[4] & 430.051 & 49.816\footnotemark[5] & 1.613\footnotemark[6] \\
	& 10 / 15               & 0    & 310.757 & 4.700    & 21.059   & 49.816 & 1.613 \\
	& 100 / 151             & 0.0  & 236.876 & 13.785   & 21.508   & 37.524 & 1.418 \\
	& 1,000 / 1,241         & 0.2  & 70.879  & 6.966    & 7.491    & 5.154  & 0.144 \\
	& 5,000 / 9,977         & 1.2  & 14.531  & 1.033(0) & 1.050(0) & 1.489  & 0.029 \\
	& 10,000 / 19,957       & 2.2  & 12.550  & 1.039(0) & 1.050(0) & 1.156  & 0.029 \\
	& 50,000 / 79,866       & 4.6  & 9.025   & 0.764(1) & 0.770(1) & 0.764  & 0.022 \\
	& 100,000 / 159,668     & 7.6  & 7.495   & 0.580(1) & 0.584(1) & 0.584  & 0.013 \\
	& 500,000 / 639,593     & 18.0 & 4.391   & 0.276(0) & 0.277(0) & 0.294  & 0.003 \\
	& 1,000,000 / 1,278,976 & 22.0 & 3.682   & 0.238(0) & 0.239(0) & 0.259  & 0.003 \\
	& 5,000,000 / 5,099,863 & 46.1 & 0.675   & 0.041(1) & 0.041(1) & 0.009  & $-0.000$ \\
\end{tabular}
\end{ruledtabular}

\footnotetext[1]{
\setlength{\baselineskip}{1em}
For each internuclear separation $R$, the $E_{\text{var}}$, $E_{\text{var}}+\Delta E^{(2)}$, and
$E_{\text{var}}+\Delta E_\text{r}^{(2)}$ energies are reported as errors, in millihartree,
relative to the extrapolated $E_{\text{var}}+\Delta E_{\text{r}}^{(2)}$ energy found using
a linear fit based on the last four $E_{\text{var},k}+\Delta E_{\text{r},k}^{(2)}$ values
leading to the largest CIPSI wave function obtained with $N_{\text{det(in)}} = 5,000,000$,
plotted against the corresponding $\Delta E_{\text{r},k}^{(2)}$ corrections,
following the procedure used in Ref.\ \onlinecite{cipsi_benzene}.
These extrapolated $E_{\text{var}}+\Delta E_{\text{r}}^{(2)}$ energies at $R = R_e$, $1.5 R_e$, $2 R_e$, and $5 R_e$
are $-199.104422(6)$, $-199.069043(1)$, $-199.060152(8)$, and $-199.059647(11)$ hartree, respectively,
where the error bounds in parentheses correspond to the uncertainty associated with the linear fit.
The error bounds for the $E_{\text{var}}+\Delta E^{(2)}$ and $E_{\text{var}}+\Delta E_\text{r}^{(2)}$ energies
obtained at the various values of $N_{\text{det(in)}}$ reflect on the semi-stochastic design of the
$\mathcal{V}_{\text{ext}}^{(k)}$ spaces discussed in the main text, but they ignore the uncertainties characterizing the
reference $E_{\text{var}}+\Delta E_{\text{r}}^{(2)}$ energies obtained in the above extrapolation procedure.
}
\footnotetext[2]{
\setlength{\baselineskip}{1em}
The CC($P$) and CC($P$;$Q$) energies are reported as errors relative 
to CCSDT, in millihartree. The total CCSDT energies at $R = R_e$, $1.5 R_e$, $2 R_e$, and $5 R_e$
are $-199.102796$, $-199.065882$, $-199.058201$, and $-199.058586$ hartree, respectively.}
\footnotetext[3]{
\setlength{\baselineskip}{1em}
Equivalent to RHF.}
\footnotetext[4]{
\setlength{\baselineskip}{1em}
Equivalent to the second-order MBPT energy using the Epstein--Nesbet denominator.}
\footnotetext[5]{
\setlength{\baselineskip}{1em}
Equivalent to CCSD.}
\footnotetext[6]{
\setlength{\baselineskip}{1em}
Equivalent to CR-CC(2,3).}
\end{table*}


\squeezetable
\begin{table*}[h!]
\caption{\label{tab:table3}
Convergence of the CC($P$) and CC($P$;$Q$) energies toward CCSDT, alongside the variational and
perturbatively corrected CIPSI energies, for the $\mathrm{F_2}$/cc-pVTZ molecule in which the
F--F bond length $R$ was fixed at $2 R_e$, where $R_e = 2.66816$ bohr is the equilibrium geometry.
For each value of the wave function termination parameter $N_{\text{det(in)}}$,
the $P$ space used in the CC($P$) calculations consisted of all singles and doubles
and a subset of triples contained in the final $|\Psi^{(\text{CIPSI})}\rangle$ state of
the underlying CIPSI run, whereas the $Q$ space needed to compute the corresponding CC($P$;$Q$)
correction $\delta(P;Q)$ was defined as the remaining triples absent in $|\Psi^{(\text{CIPSI})}\rangle$.
In all post-RHF calculations, the two lowest-lying core orbitals were frozen and the spherical components
of $d$ and $f$ orbitals were employed throughout. Each CIPSI run was initiated from the RHF reference
determinant and the MBPT-based stopping parameter $\eta$ was set at $10^{-6}$ hartree. The input parameter
$f$ controlling the CIPSI wave function growth was set at the default value of 2.
}
\footnotesize
\begin{ruledtabular}
\begin{tabular}{p{3.5cm} c c c c c c}
\textrm{$N_{\text{det(in)}}$ / $N_{\text{det(out)}}$}&
\textrm{\% of triples}&
\textrm{$E_{\text{var}}$\footnotemark[1]}&
\textrm{$E_{\text{var}}+\Delta E^{(2)}$\footnotemark[1]}&
\textrm{$E_{\text{var}}+\Delta E^{(2)}_{\text{r}}$\footnotemark[1]}&
\textrm{CC$(P)$\footnotemark[2]}&
\textrm{CC$(P;Q)$\footnotemark[2]} \\
\colrule
        1 / 1                 & 0 & 758.849\footnotemark[3] & $-165.740$\footnotemark[4] & 340.460 & 62.819\footnotemark[5] & 4.254\footnotemark[6] \\
        10 / 18               & 0    & 441.567 & $-0.554$    & 31.337    & 62.819 & 4.254 \\
        100 / 156             & 0.00 & 393.749 & 6.420     & 28.790    & 58.891 & 3.683 \\
        1,000 / 1,277         & 0.01 & 253.172 & 13.595(0) & 20.323    & 42.564 & 1.579 \\
        5,000 / 5,118         & 0.03 & 123.591 & 10.874(1) & 12.149(1) & 18.036 & 0.345 \\
        10,000 / 10,239       & 0.06 & 73.122  & 7.202(5)  & 7.636(5)  & 11.439 & 0.198 \\
        50,000 / 82,001       & 0.84 & 29.674  & 3.371(2)  & 3.428(2)  & 4.898  & 0.061 \\
        100,000 / 163,866     & 1.58 & 27.002  & 3.068(2)  & 3.113(2)  & 4.157  & 0.049 \\
        500,000 / 655,859     & 3.75 & 22.301  & 2.517(1)  & 2.547(1)  & 3.111  & 0.014 \\
        1,000,000 / 1,311,633 & 5.58 & 20.244  & 2.292(1)  & 2.316(1)  & 2.739  & 0.009 \\
        5,000,000 / 5,253,775 & 13.3 & 14.499  & 1.645(1)  & 1.657(1)  & 1.866  & $-0.015$
\end{tabular}
\end{ruledtabular}

\footnotetext[1]{
\setlength{\baselineskip}{1em}
The $E_{\text{var}}$, $E_{\text{var}}+\Delta E^{(2)}$, and
$E_{\text{var}}+\Delta E_\text{r}^{(2)}$ energies are reported as errors, in millihartree,
relative to the extrapolated $E_{\text{var}}+\Delta E_{\text{r}}^{(2)}$ energy found using
a linear fit based on the last four $E_{\text{var},k}+\Delta E_{\text{r},k}^{(2)}$ values
leading to the largest CIPSI wave function obtained with $N_{\text{det(in)}} = 5,000,000$,
plotted against the corresponding $\Delta E_{\text{r},k}^{(2)}$ corrections,
following the procedure used in Ref.\ \onlinecite{cipsi_benzene}.
The extrapolated $E_{\text{var}}+\Delta E_{\text{r}}^{(2)}$ energy is $-199.242119(59)$ hartree,
where the error bounds in parentheses correspond to the uncertainty associated with the linear fit.
The error bounds for the $E_{\text{var}}+\Delta E^{(2)}$ and $E_{\text{var}}+\Delta E_\text{r}^{(2)}$ energies
obtained at the various values of $N_{\text{det(in)}}$ reflect on the semi-stochastic design of the
$\mathcal{V}_{\text{ext}}^{(k)}$ spaces discussed in the main text, but they ignore the uncertainties characterizing the
reference $E_{\text{var}}+\Delta E_{\text{r}}^{(2)}$ energy obtained in the above extrapolation procedure.
}
\footnotetext[2]{
\setlength{\baselineskip}{1em}
The CC($P$) and CC($P$;$Q$) energies are reported as errors relative to
CCSDT, in millihartree. The total CCSDT energy is $-199.238344$ hartree.}
\footnotetext[3]{
\setlength{\baselineskip}{1em}
Equivalent to RHF.}
\footnotetext[4]{
\setlength{\baselineskip}{1em}
Equivalent to the second-order MBPT energy using the Epstein--Nesbet denominator.}
\footnotetext[5]{
\setlength{\baselineskip}{1em}
Equivalent to CCSD.}
\footnotetext[6]{
\setlength{\baselineskip}{1em}
Equivalent to CR-CC(2,3).}
\end{table*}


\squeezetable
\begin{table*}[h!]
\caption{\label{tab:table4}
Convergence of the CC($P$) and CC($P$;$Q$) energies toward CCSDT, alongside the variational
and perturbatively corrected CIPSI energies, for the reactant (R) and
transition-state (TS) species involved in the automerization of cyclobutadiene, as described
by the cc-pVDZ basis set, and for the corresponding barrier height. The R and TS geometries,
optimized using the MR-AQCC approach, were taken from Ref.\ \onlinecite{MR-AQCC}.
For each value of the wave function termination parameter $N_{\text{det(in)}}$,
the $P$ space used in the CC($P$) calculations consisted of all singles and doubles
and a subset of triples contained in the final $|\Psi^{(\text{CIPSI})}\rangle$ state of
the underlying CIPSI run, whereas the $Q$ space needed to compute the corresponding CC($P$;$Q$)
correction $\delta(P;Q)$ was defined as the remaining triples absent
in $|\Psi^{(\text{CIPSI})}\rangle$. In all post-RHF calculations, the four lowest-lying
core orbitals were frozen and the spherical components of $d$ orbitals were employed throughout. 
Each CIPSI run was initiated from the RHF reference determinant and the MBPT-based stopping
parameter $\eta$ was set at $10^{-6}$ hartree. The input parameter $f$ controlling the
CIPSI wave function growth was set at the default value of 2.
}
\footnotesize
\begin{ruledtabular}
\begin{tabular}{l p{5.2cm} c c c c c c}
\textrm{Species}&
\textrm{$N_{\text{det(in)}}$ / $N_{\text{det(out)}}$}&
\textrm{\% of triples}&
\textrm{$E_{\text{var}}$\footnotemark[1]}&
\textrm{$E_{\text{var}}+\Delta E^{(2)}$\footnotemark[1]}&
\textrm{$E_{\text{var}}+\Delta E^{(2)}_{\text{r}}$\footnotemark[1]}&
\textrm{CC$(P)$\footnotemark[2]}&
\textrm{CC$(P;Q)$\footnotemark[2]} \\
\colrule
	R & 1 / 1 & 0 & 598.120\footnotemark[3] & $-83.736$\footnotemark[4] & 120.809 & 26.827\footnotemark[5] & 0.848\footnotemark[6] \\
	& 50,000 / 55,653         & 0.0  & 121.880 & 26.065(182) & 28.096(178) & 25.468 & 0.678 \\
	& 100,000 / 111,321       & 0.1  & 109.688 & 23.819(163) & 25.397(160) & 22.132 & 0.382 \\
	& 500,000 / 890,582       & 1.2  & 93.413  & 19.049(141) & 20.167(139) & 16.260 & 0.267 \\
	& 1,000,000 / 1,781,910   & 2.0  & 89.989  & 18.322(137) & 19.348(135) & 15.359 & 0.251 \\
	& 5,000,000 / 7,125,208   & 7.9  & 78.122  & 16.311(123) & 17.045(122) & 10.794 & 0.150 \\
	& 10,000,000 / 14,253,131 & 11.8 & 73.250  & 15.514(115) & 16.146(114) & 9.632  & 0.127 \\
	& 15,000,000 / 28,493,873 & 25.8 & 60.872  & 12.842(96)  & 13.260(95)  & 4.817  & 0.046 \\
        &                         &      &         &             &             &        &       \\ [-1.5mm]
	TS & 1 / 1 & 0 & 632.707\footnotemark[3] & $-102.816$\footnotemark[4] &  282.246 & 47.979\footnotemark[5] & 14.636\footnotemark[6] \\
	& 50,000 / 56,225         & 0.0  & 146.895 & 45.357(180) & 47.696(176) & 42.132 & 9.563 \\
	& 100,000 / 112,481       & 0.1  & 130.832 & 36.716(183) & 38.673(179) & 31.723 & 3.507 \\
	& 500,000 / 899,770       & 1.0  & 93.288  & 18.106(139) & 19.251(137) & 14.742 & 0.432 \\
	& 1,000,000 / 1,800,183   & 1.6  & 89.049  & 17.458(142) & 18.482(140) & 13.645 & 0.412 \\
	& 5,000,000 / 7,195,780   & 5.4  & 78.472  & 15.587(124) & 16.346(123) & 10.720 & 0.260 \\
	& 10,000,000 / 14,400,744 & 9.7  & 71.784  & 14.397(114) & 15.016(113) & 8.358  & 0.155 \\
	& 15,000,000 / 28,793,512 & 15.2 & 63.375  & 12.587(102) & 13.058(101) & 7.080  & 0.108 \\
        &                         &      &         &             &             &        &       \\ [-1.5mm]
	Barrier & 1 / 1 ; 1  & 0 ; 0 & 21.703\footnotemark[3] & $-11.973$\footnotemark[4] & 101.303 & 13.274\footnotemark[5] & 8.653\footnotemark[6] \\
	& 50,000 / 55,653 ; 56,225             & 0.0 ; 0.0   & 15.697 & 12.106(161) & 12.299(157) & 10.457 & 5.576 \\
	& 100,000 / 111,321 ; 112,481            & 0.1 ; 0.1   & 13.268 & 8.093(154)  & 8.331(151)  & 6.018  & 1.961 \\
	& 500,000 / 890,582 ; 899,770            & 1.2 ; 1.0   & $-0.079$ & $-0.592(124)$ & $-0.574(122)$ & $-0.953$ & 0.104 \\
	& 1,000,000 / 1,781,910 ; 1,800,183        & 2.0 ; 1.6   & $-0.590$ & $-0.542(124)$ & $-0.544(122)$ & $-1.075$ & 0.101 \\
	& 5,000,000 / 7,125,208 ; 7,195,780        & 7.9 ; 5.4   & 0.220  &   $-0.454(110)$ & $-0.439(109)$ & $-0.047$ & 0.069 \\
	& 10,000,000 / 14,253,131 ; 14,400,744     & 11.8 ; 9.7  & $-0.920$ & $-0.701(102)$ & $-0.710(100)$ & $-0.800$ & 0.017 \\
	& 15,000,000 / 28,493,873 ; 28,793,512 & 25.8 ; 15.2 & 1.571  & $-0.159(88)$  & $-0.127(87)$  & 1.420  & 0.039 \\
\end{tabular}
\end{ruledtabular}

\footnotetext[1]{
\setlength{\baselineskip}{1em}
For each of the two species, the $E_{\text{var}}$, $E_{\text{var}}+\Delta E^{(2)}$, and
$E_{\text{var}}+\Delta E_\text{r}^{(2)}$ energies are reported as errors, in millihartree,
relative to the extrapolated $E_{\text{var}}+\Delta E_{\text{r}}^{(2)}$ energy found using
a linear fit based on the last four $E_{\text{var},k}+\Delta E_{\text{r},k}^{(2)}$ values
leading to the largest CIPSI wave function obtained with $N_{\text{det(in)}} = 15,000,000$,
plotted against the corresponding $\Delta E_{\text{r},k}^{(2)}$ corrections,
following the procedure used in Ref.\ \onlinecite{cipsi_benzene}.
These extrapolated $E_{\text{var}}+\Delta E_{\text{r}}^{(2)}$ energies for the R and TS species 
are $-154.249292(314)$ and $-154.235342(321)$ hartree, respectively,
where the error bounds in parentheses correspond to the uncertainty associated with the linear fit.
The error bounds for the $E_{\text{var}}+\Delta E^{(2)}$ and $E_{\text{var}}+\Delta E_\text{r}^{(2)}$ energies
obtained at the various values of $N_{\text{det(in)}}$ reflect on the semi-stochastic design of the
$\mathcal{V}_{\text{ext}}^{(k)}$ spaces discussed in the main text, but they ignore the uncertainties characterizing the
reference $E_{\text{var}}+\Delta E_{\text{r}}^{(2)}$ energies obtained in the above extrapolation procedure.
The $E_{\text{var}}$, $E_{\text{var}}+\Delta E^{(2)}$, and $E_{\text{var}}+\Delta E_\text{r}^{(2)}$
barrier heights are reported as errors, in kcal/mol, relative to the reference value of 8.753(0) kcal/mol
obtained using the extrapolated $E_{\text{var}}+\Delta E_{\text{r}}^{(2)}$ energies of the R and TS species.
}
\footnotetext[2]{
\setlength{\baselineskip}{1em}
The CC($P$) and CC($P$;$Q$) energies of the R and TS species are reported
as errors relative to CCSDT, in millihartree. The total CCSDT energies of the R and TS 
species are $-154.244157$ and $-154.232002$ hartree, respectively. The CC($P$) and CC($P$;$Q$)
barrier heights are reported in kcal/mol relative to the CCSDT value of 7.627 kcal/mol.}
\footnotetext[3]{
\setlength{\baselineskip}{1em}
Equivalent to RHF.}
\footnotetext[4]{
\setlength{\baselineskip}{1em}
Equivalent to the second-order MBPT energy using the Epstein--Nesbet denominator.}
\footnotetext[5]{
\setlength{\baselineskip}{1em}
Equivalent to CCSD.}
\footnotetext[6]{
\setlength{\baselineskip}{1em}
Equivalent to CR-CC(2,3).}
\end{table*}


\end{document}